\documentclass[twocolumn]{aastex701}

\usepackage{textcomp}
\usepackage{subcaption}
\usepackage[flushleft]{threeparttable}
\usepackage{amsmath}
\usepackage[T1]{fontenc}

\accepted{September 29, 2026}
\submitjournal{ApJ}

\begin{document}

\title{Tentative Detection of FeS toward Orion SrcI}

\author[orcid=0009-0002-3398-4627]{A.\ Tasa-Chaveli}
\affiliation{Centro de Astrobiolog\'ia (CAB), CSIC-INTA, Ctra.\ de Ajalvir, km 4, Torrej\'on de Ardoz, 28850 Madrid, Spain}
\affiliation{Departamento de F\'isica de la Tierra y Astrof\'isica, Facultad CC. F\'isicas, Universidad Complutense de Madrid, Spain}
\email[show]{atasa@cab.inta-csic.es}

\author[orcid=0000-0002-0183-8927]{H.~S.~P.\ M\"uller}
\affiliation{Astrophysik/I.\ Physikalisches Institut, Universit\"at zu K\"oln, Z\"ulpicher Str.\ 77, 50937 K\"oln, Germany}
\email{hspm@ph1.uni-koeln.de}

\author[orcid=0000-0002-3078-9482]{\'A.\ S\'anchez-Monge}
\affiliation{Institut de Ci\`encies de l’Espai (ICE), CSIC, Campus UAB, Carrer de Can Magrans s/n, 08193 Bellaterra, Barcelona, Spain}
\affiliation{Institut d'Estudis Espacials de Catalunya (IEEC), 08860 Castelldefels, Barcelona, Spain}
\email[show]{asanchez@ice.csic.es}

\author[orcid=0000-0001-6317-6343]{A.\ Fuente}
\affiliation{Centro de Astrobiolog\'ia (CAB), CSIC-INTA, Ctra.\ de Ajalvir, km 4, Torrej\'on de Ardoz, 28850 Madrid, Spain}
\email{afuente@cab.inta-csic.es}

\author[orcid=0000-0001-6431-9633]{A.\ Ginsburg}
\affiliation{Department of Astronomy, University of Florida, P.O.\ Box 112055, Gainesville, FL 32611, USA}
\email{adam.g.ginsburg@gmail.com}

\author[orcid=0000-0002-9277-8025]{Th.\ M\"oller}
\affiliation{Astrophysik/I.\ Physikalisches Institut, Universit\"at zu K\"oln, Z\"ulpicher Str.\ 77, 50937 K\"oln, Germany}
\email{moeller@ph1.uni-koeln.de}

\begin{abstract}
Iron sulfide (FeS) is a key refractory compound in astrophysical environments. It is commonly found in meteorites and evidence of FeS has only been reported in solid form in a protoplanetary environment. Furthermore, FeS is of astrobiological interest since it acts as a precursor to the formation of  pyrite (FeS$_2$), which plays a key role in anaerobic metabolism. We present millimeter-wave spectroscopic calculations for gaseous FeS and report its tentative detection toward the massive disk Orion SrcI. If confirmed, this would constitute the first ever detection of gaseous FeS in the interstellar medium. Our data are consistent with a FeS column density of $\rm \sim1.7 \times 10^{15}~cm^{-2}$. Comparing to other sulfur reservoirs, the amount of sulfur locked in FeS is  more than three orders of magnitude lower than that contained in SO$_2$ and SiS. However, FeS emission may arise from a different region as it is the case for the metallic salt NaCl. Our results support the presence of FeS in interstellar grains and are consistent with grain destruction playing an important role in the origin of the FeS emission. The observations, however, are still insufficient to discern whether FeS is a main sulfur reservoir in molecular clouds or a primary carrier of refractory iron-bearing material.
\end{abstract}


%
\section{Introduction}\label{sec:introduction}

The formation of iron- and sulfur-bearing species in protoplanetary disks plays a key role in the chemical evolution of refractory materials, which are the building blocks of planet formation. Refractory species are compounds with high sublimation temperatures, typically of several hundred kelvin. Detecting and studying these species allow us to investigate unexplored environments and physical conditions, such as those found in the innermost regions of protoplanetary disks and massive protostellar sources.

Iron sulfide (FeS) is a key  compound in astrochemistry and astrobiology. First, FeS acts as a precursor to the formation of more complex S-bearing species such as pyrite (FeS$_2$), which is one of the most abundant sulfide minerals on Earth \citep{Liu2025}. In this context, iron sulfides are of great astrobiological interest, as they can drive chemical reactions involved in prebiotic processes, with pyrite thought to have a leading role in the metabolism of primitive life \citep{Wachtershauser1988}.
Secondly, current dust evolution models face a long-standing discrepancy between dust production in stellar sources and its rapid destruction in the interstellar medium (ISM), suggesting that a significant fraction of dust must grow within the ISM \citep[see e.g.,][]{Dwek2016}. Refractory compounds containing abundant elements such as iron are therefore expected to contribute to this grain growth. 

In addition, FeS links both iron and sulfur chemistry, two of the most abundant elements in the Universe \citep[Fe/H $\approx3.3 \times10^{-5}$ and S/H $\approx1.5 \times10^{-5}$,][]{Asplund2009}. Iron is mainly formed in Type~Ia and core collapse supernovae \citep{Thielemann1986}. As a result, a substantial fraction ($\sim$65\%) of iron is injected into the ISM in gaseous form \citep{Dwek2016}. However, only $<10\%$ of the expected iron is observed in the gas phase in warm and cold regions \citep{Jenkins2009}, indicating a strong depletion of Fe compared to its elemental abundance. Using deep Chandra/HETG spectra of the bright X-ray binary GX 340+00, \citet{Rogantini2026} estimated that $\sim$8\% of Fe might be associated with sulfides. Moreover, sulfur is another key element that has remained a challenge for decades: the observed gas-phase abundances of S-bearing molecules in the dense gas constitute only $<1\%$ of the expected amount \citep[e.g.,][]{Ruffle1999, Fuente2023}, and refractory species have been proposed as an important sulfur reservoir.

Despite its relevance, the FeS chemistry remains poorly characterized. Although it is commonly found in meteorites \citep{Cook2017}, FeS has not yet been detected in the gas phase in the ISM thus far. Evidence for FeS has only been reported in solid form in protoplanetary environments  \citep[e.g.,][]{Keller2002}.
As a refractory species, its detection in the gas phase is particularly challenging, as its release from dust grains requires extreme conditions, such as high temperatures \citep[$>500$~K;][]{Kobayashi2011}, intense radiation, or strong shocks. This has limited their search in the interstellar medium and star-forming regions. 

Recent high-angular-resolution observations toward the protostellar disk of Orion~Src\,I led to the detection of the refractory molecules NaCl and KCl in the ISM \citep{Ginsburg2019}. This opened a new door to search for other refractory species. Follow-up studies have reveled NaCl and KCl in multiple disks, together with AlO \citep{Tanaka2020, Ginsburg2023}.
More recently, TiO has also been detected toward Orion SrcI \citep{Zhang2026-TiO}. Independently, the sulfur-bearing compounds MgS and NaS have been detected toward the Galactic Center by \citet{ReyMontejo2024}, while \citet{Tasa-Chaveli2025} detected CaS in the massive disk G351.77-mm1. 

This letter presents the spectroscopic properties of FeS and the search for this species in the disk around the massive protostellar object Orion Source I (hereafter SrcI), based on high spatial resolution observations carried out with the Atacama Large Millimeter/submillimeter Array (ALMA).

%
\section{Spectroscopic Properties of \texorpdfstring{F\lowercase{e}S}{FeS}}\label{sec:lab-spectroscopic}

The FeS molecule is, as the isovalent FeO, a radical with four unpaired electrons forming a $X^{5}\Delta _{i}$ electronic ground state with the $\Omega = 4$ spin ladder being lowest in energy and $\Omega = 0$ about 360~cm$^{-1}$ or 520~K above $\Omega = 4$ \citep[calculated from the spectroscopic parameters of][]{FeS_rot_2004} being highest. 
Whereas the rotational spectrum of FeO was characterized quite extensively \citep{FeO_rot_1984,FeO_rot_1987,FeO_rot_1996,FeO_rot_2020}, there is only one report by \citet{FeS_rot_2004} dealing with the rotational spectrum of FeS between 219 and 395~GHz. 
They reported transition frequencies in the lowest four spin ladders $\Omega = 4$ to $\Omega = 1$ pertaining to the $v = 0$ ground vibrational state, $\Omega = 4$ and 3 transition frequencies in its $v = 1$ first excited vibrational state, as well as data that were assigned tentatively to transitions within the first excited $A^{5}\Sigma ^{+}$ electronic state. 
The $v = 0$ $\Omega = 4$, 3, and 1 data were reproduced well by the spectroscopic model \citep{FeS_rot_2004}, including the resolved $\Lambda$-doubling splitting in $\Omega = 1$, whereas the $\Lambda$-doubling splitting and the positions of the $\Omega = 2$ lines deviated substantially; this deviation was interpreted as a perturbation from an as of yet unidentified perturber. 
We point out that no $\Lambda$-doubling was resolved for $\Omega = 4$ and 3 in the rotational spectrum of FeS, similar to FeO, for which $\Lambda$-doubling remained unresolved even in $\Omega = 2$ \citep{FeO_rot_1996}. 
The dipole moment of FeS was calculated to be 5.41~D \citep{FeS_MRCI_2022}. 

The first excited vibrational state was measured to be $\sim$518~cm$^{-1}$ or 745~K above $v = 0$ \citep{FeS_vib_2011}. 
The transitions within this vibrational state may be observable alongside those of $v = 0$ in the case of high vibrational temperatures in an astronomical source. 
\citet{FeS_rot_2004} were only able to identify transitions of $\Omega = 4$ and 3 in $v = 1$ which does not permit simultaneous fitting of the spin-orbit parameter $A_{\rm SO}$ and the spin-spin coupling parameter $\lambda$. 
They fixed $\lambda$ and the distortion corrections $\lambda _D$ and $A_D$ to zero and obtained an $A_{\rm SO}$ value much larger in magnitude than the ground state value, $-$74.9~~cm$^{-1}$ versus $-$44.7~cm$^{-1}$. 
Assuming the $v = 1$ fine structure parameters might be somewhat closer to those in $v = 0$, we fit $A_{\rm SO}$ with the remaining three fine structure parameters kept fixed to their $v = 0$ counterparts. 
The resulting $v = 1$ $A_{\rm SO}$ is only 2\% smaller in magnitude than the respective $v = 0$ value. 
While it is unclear if our $v = 1$ model is better than the published one, the assumption of the fine structure parameters being quite similar in $v = 1$ and 0 appears to be more reasonable until proven to be wrong.

The first excited $A^{5}\Sigma ^{+}$ electronic state, on the other hand, was determined to be 3952~cm$^{-1}$ or about 5700~K above the ground vibronic state \citep{FeS_A-egy_1996}, deemed to be too high in energy for our present astronomical data.

Pickett's spfit and spcat programs \citep{spfit_1991} were employed for fitting and calculation of the spectra. 
While the programs were written to model spectra conforming to Hund's case (b), it has been shown that they may well be used for molecules better described by Hund's case (a), such as NO \citep{NO_rot_2015}, CH$_3$O \citep{CH3O_rot_2025}, and even IO \citep{IO_rot_2001}. 
The main disadvantage of using Hund's case (b) quantum numbers in such cases is that the rotational quantum number $N$ is not a good quantum number, meaning that the correlation of $N$ and $J$ to $\Omega$ and $J$ is not straightforward and often switches within a particular $\Omega$, as pointed out, for example, in the case of NO \citep{NO_rot_2015}.

Catalog files of FeS were produced for $\Omega = 4$ and 3 of $v = 0$ and 1 up to 500~GHz. 
We refrain from providing transition frequencies for $\Omega = 2$ and 0 because of its perturbation in the first case and because it was not observed at all in the second case. 
Unfortunately, spfit and spcat are only capable of treating $\Lambda$-doubling for $\Omega = 4$ to 2. Therefore, we do not present calculations for $\Omega = 1$ for now. 
The experimental $\Omega = 2$ and 1 data could be treated separately, but such calculations need to be viewed with very great caution. 

The experimental data in $v = 0$ cover 230 to 389~GHz while $v = 1$ data were taken between 325 and 387~GHz. 
The $v = 0$ calculations should be sufficiently reliable throughout, although some caution is advised for the lowest frequency transitions because effects of the spin-spin coupling increase with decreasing $J$. 
More caution is necessary in case of the $v = 1$ calculations because of the more limited and in part less accurate experimental data and because of the assumption on the spectroscopic model. 

The spacing between $\Omega$ ladders changes only slightly with $J$, such that associating transitions with given $N$ and $J$ is straightforward based on the lower state energies provided in an spcat catalog file.
In the case of FeS, $N = J - 2$ for $J \leq 5$ and $N = J - 1$ for $J \geq 6$ correlate with $\Omega = 4$ whereas $N = J$ for $J \leq 5$ and $N = J - 2$ for $J \geq 6$ correlate with $\Omega = 3$.
We provide catalog files for the individual vibrational states and $\Omega$ ladders to be able to treat them separately, but also to avoid confusion with the correlations.

Complete lists of calculated millimeter frequencies are available in the catalog section\footnote{https://cdms.astro.uni-koeln.de/classic/entries/} of the Cologne Database for Molecular Spectroscopy \citep[CDMS;][]{Mueller2005}. 
In Table~\ref{tab:FeS}, we list the frequencies of the lines discussed in this work.

\section{Search for \texorpdfstring{F\lowercase{e}S}{FeS}}\label{sec:search}

We searched for FeS using the stacked spectrum toward SrcI reported by \citet{Ginsburg2018}.  This spectrum was obtained using Band 6 data from the  ALMA project 2016.1.00165.S (PI: J. Bally). 
The spectral cubes were generated with robust 0.5 weighting, with a synthesized beam of  $\approx 0\farcs04\times 0\farcs03$. The velocity field was derived from the NaCl $v = 1 \ J=18-17$ line \citep[][]{Ginsburg2019}, and the spectra were shifted to 0~km~s$^{-1}$ and averaged to produce stacked spectra over the NaCl-emitting region \citep[$\approx$ 0.03 arcsec$^2$, corresponding to about 20 beams; see Appendix D of][for a more detailed description]{Ginsburg2018}.
 The NaCl emission is thought to trace the innermost and hottest gas, likely associated with the shocked regions, where complex organic molecules (COMs) may be efficiently destroyed. This optimizes the possibility of detecting FeS while minimizing possible contamination from COM lines.

Figure~\ref{fig:total-spectrum} shows the Band 6 stacked spectrum toward the disk SrcI (with coordinates $\alpha(\mathrm{J2000})=05^\mathrm{h}35^\mathrm{m}14.520^\mathrm{s}$, $\delta(\mathrm{J2000})=-05^{\circ}22^{\prime}30.623^{\prime\prime}$), including all species reported by \citet{Ginsburg2019}, together with newly identified SiS \citep[frequencies derived from][]{SiS_rot_2007} and SO$_2$ lines \citep[frequencies derived from][]{SO2_rot_2005}. The labeled transitions are listed in Table~\ref{tab:all_lines}. 
The rms noise level of the stacked spectrum at this band is 1.1~K and the channel width $\sim$1.3~km~s$^{-1}$. 
Note that these observations, together with those in Bands 3 and 7 \citep[see][]{Ginsburg2019} show no bright emission from COMs.

%
\subsection*{Tentative detection of \texorpdfstring{F\lowercase{e}S}{FeS}}\label{sec:results}
\begin{figure*}
    \begin{subfigure}{\linewidth}
    \centering
    \includegraphics[width=\linewidth]{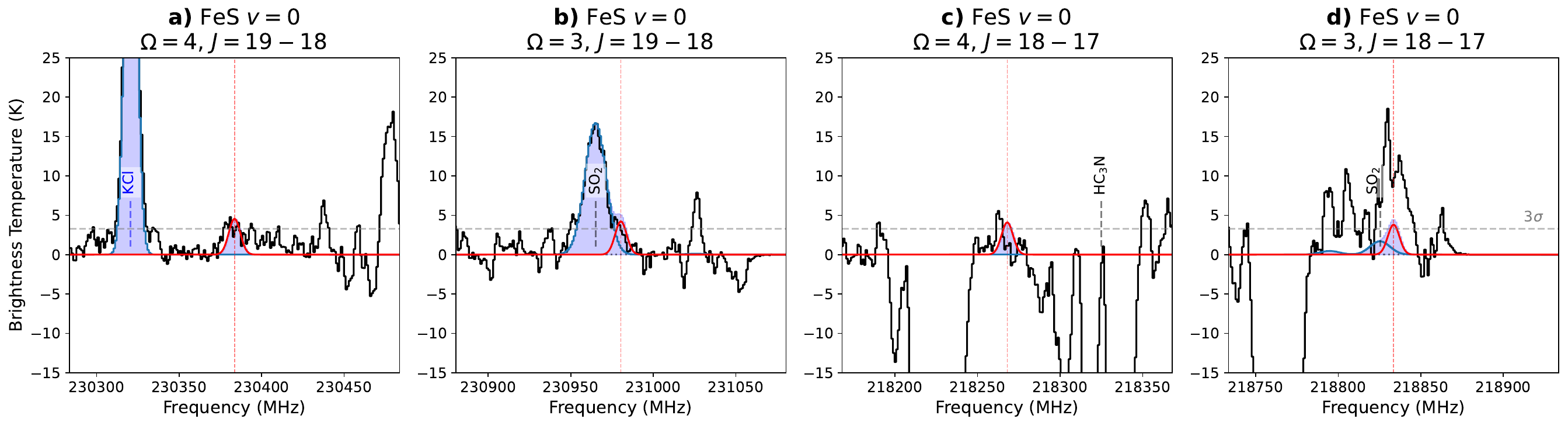}
    \end{subfigure}
    \begin{subfigure}{\linewidth}
    \centering
    \includegraphics[width=\linewidth]{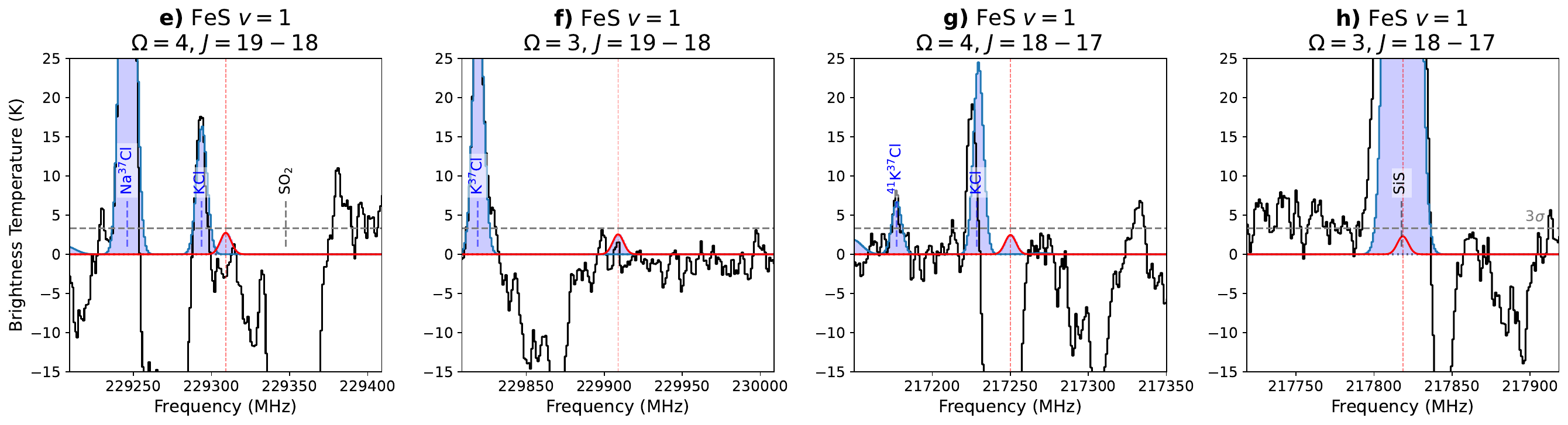}
    \end{subfigure}
\caption{Zoom-in panels around the FeS ground-state (upper panels) and $v=1$ (bottom panels) transitions included in the spectra of Orion SrcI. The observed data are shown with black histograms. The synthetic spectrum generated excluding FeS is shown with a blue line, while the best-fit model of FeS is shown in red. The blue shaded region corresponds to the combined contribution of FeS and the remaining species. The horizontal gray dashed line indicates the 3$\sigma$ level. It should be noticed that the $v=0 \ J = 19 \rightarrow 18$ lines cover 8 - 9 channels resulting in detections with SNR > 8. }
\label{fig:FeS-zooms}
\end{figure*}
The observed frequency range covers four ground-state and four $v=1$ transitions of iron sulfide (see Fig.~\ref{fig:FeS-zooms}). Each transition is split into two unresolved components, resulting in 16 spectroscopic entries listed in Table~\ref{tab:FeS}. 
We clearly identify two $J=19-18$ ground-state transitions (see panels~a and b of Fig.~\ref{fig:FeS-zooms}). The transition in panel a is fairly clean, while the transition in panel b is only weakly blended with a SO$_2$ line and remains distinguishable.
The $J=18-17$ $\Omega=4$ might be affected by absorptions (panel~c) and the $J=18-17$ $\Omega=3$ appears heavily contaminated with other species (panel~d). The vibrationally excited $v=1$ transitions are all blended with absorption features or intense identified compounds (see bottom panels of Fig.~\ref{fig:FeS-zooms}). 

The FeS lines were modeled by computing synthetic spectra assuming LTE conditions, based on the spectroscopic parameters derived in this work (listed in Table~\ref{tab:FeS}) and adopting Gaussian line profiles. A linewidth of 11~km~s$^{-1}$ provides the best fit to the observed FeS $v=0$ transitions, consistent with the values obtained for the other refractory species detected in this source (see Sect.~\ref{sec:discussion}). We note that the velocity offset was set to 0~km~s$^{-1}$, although a shift to $\approx1$ would improve the fit.
We assume a rotational temperature of 1500~K, based on our H$_2$O $v_2=1$ and $^{29}$SiO $v=0$ analysis (see Sect.~\ref{sec:discussion}). 
For this temperature, the best-fit model corresponds to a FeS column density of $\rm 1.7\times10^{15}~cm^{-2}$. To assess the uncertainty associated with the adopted rotational temperature, we considered a range of temperatures between 500 and 1500~K (see Table~\ref{tab:results}). The derived column density varies by a factor of 3 over the explored temperature range. The red solid line in Fig.~\ref{fig:FeS-zooms} shows the best-fit model for FeS. The synthetic spectrum reproduces the observed $v=0$ transitions while remaining consistent with the non-detection of the $v=1$ transitions, whose predicted emission is below the 3$\sigma$ level.

In order to ensure that the two  fairly clean lines shown in panels a and b correspond to FeS transitions, we examined the spectral vicinity ($\rm \pm~ FWHM/2  \approx \pm4.2$~MHz) of each transition and searched for any other potential molecular species in the commonly-used Cologne Database for Molecular Spectroscopy \citep[CDMS;][]{Mueller2005} and Jet Propulsion Laboratory (JPL) line catalogs \citep{Pickett1998}. All possible contaminants are listed in Table~\ref{tab:blending}.

For each potential contaminant, a synthetic spectrum was generated using the eXtended CASA Line Analysis Software Suite \citep[XCLASS;][]{Moeller2017}, which considers LTE excitation conditions. 
We explored temperatures ranging from 500 to 1500~K, fixing the source size to $0\farcs1$ based on the moment-0 maps of the resolved NaCl emission presented by \citet{Ginsburg2019} (beam-filling factor of $\approx 1$),
and the line width to 11~km~s$^{-1}$ (similar to that derived for the salts; see Table~\ref{tab:results}). Following the same methodology as \citet{Tasa-Chaveli2025}, we discarded all candidate contaminants based on line inconsistencies: if we try to fit the observed feature with any of these species, other transitions expected at different frequencies and not affected by absorption would be too bright to reproduce the observations.

Furthermore, reproducing the emission features at 230.3836~GHz and 230.9804~GHz would require an abundance $\gtrsim 10^{-8}$ for more than 50\% of the considered molecules, assuming an H$_2$ column density\footnote{$N_{\rm H_2}$ was derived from the continuum map, with an integrated flux density of $\sim$87 mJy, assuming a disk size of 0$\farcs1$, an optical depth $\kappa=0.01$~cm$^2$~g$^{-1}$ and a dust temperature $T_{\rm dust}=400$~K. Varying the temperature between 200~K and 1500~K introduces an uncertainty of a factor of two in the derived H$_2$ column density.} of $\rm 1.8\times10^{25}~cm^{-2}$. Such abundances are unrealistically high for most of the COMs and refractory compounds, further supporting their rejection. 
Moreover, the detection of COMs in emission is chemically unlikely at the high-angular-resolution scales probed here, which trace the hot, shocked inner region. Consistent with this scenario, the detected emission across the full frequency coverage reported by \citet{Ginsburg2019} is dominated by refractory compounds, SiO and high-excitation SO$_2$. 

In this regard, some refractory compounds that are potential contaminants were identified (see Table~\ref{tab:blending}) and also rejected: Si$^{33}$S $v=2$, a possible contaminant of the FeS $\Omega =4 \ 19-18$ transition, and c-SiC$_3$, which is in the spectral vicinity of FeS $\Omega =3 \ 19-18$. We generated a synthetic spectrum of Si$^{33}$S $v=2$ based on the upper limit derived for SiS $v=2$ ($N<2.6 \times 10^{16}$~cm$^{-2}$; Table~\ref{tab:results}). Assuming  the canonical isotopic ratio $\rm ^{32}S/^{33}S\sim$127 \citep{Lodders2020}, no Si$^{33}$S $v=2$ emission is expected in our spectrum. In the case of the bicyclic molecule c-SiC$_3$, this species has so far only  been detected toward the evolved carbon-rich star IRC +10216 \citep{Apponi1999-c-SiC3}, and can be ruled out based on line inconsistencies, as fitting the observed feature would predict additional transitions that are not detected in the spectrum.  

Consequently, we conclude that FeS is the most likely carrier of the observed features at 230.3836~GHz and 230.9804~GHz.

%
\begin{table*}[h!]
\begin{center}
\caption{Physical parameters of molecules (tentatively) detected toward Orion SrcI and discussed in this work\label{tab:results}}
\begin{tabular}{l r r c c c}
\tableline\tableline \noalign{\smallskip}
  Species  
& \multicolumn{1}{c}{$N_\mathrm{mol}(500~\mathrm{K})$}
& \multicolumn{1}{c}{$N_\mathrm{mol}(1500~\mathrm{K})$}
& $T_\mathrm{rot}$
& $\Delta v$
& $v_\mathrm{offset}$
\\

& \multicolumn{1}{c}{(cm$^{-2}$)}
& \multicolumn{1}{c}{(cm$^{-2}$)}
& (K)
& (km~s$^{-1}$)
& (km~s$^{-1}$)
\\
\tableline \noalign{\smallskip}
FeS $v=0$     & \phn$5.4\times10^{14}$   &  \phn$1.7\times10^{15}$ & \phn$500-1500$ & 11.0 &  0 \\
NaCl $v=1$    & \phn$7.5${\raisebox{0.5ex}{\tiny$\substack{+0.1 \\ -0.4}$}}$\times10^{15}$   &  
 \phn$2.0${\raisebox{0.5ex}{\tiny$\substack{+0.1 \\ -1.0}$}}$\times10^{16}$& \phn$500-1500$ & \phn11.7${\raisebox{0.5ex}{\tiny$\substack{+0.2 \\ -0.3}$}}$ & \phn0.1${\raisebox{0.5ex}{\tiny$\substack{+0.1 \\ -0.2}$}}$ \\   
NaCl $v=2$    &   \phn$2.0${\raisebox{0.5ex}{\tiny$\substack{+0.1 \\ -0.1}$}}$\times10^{16}$   &  
 \phn$2.8${\raisebox{0.5ex}{\tiny$\substack{+0.1 \\ -0.7}$}}$\times10^{16}$ & \phn$500-1500$ & \phn11.7 & \phn0.1  \\   
Na$^{37}$Cl $v=0$    &   \phn$2.3${\raisebox{0.5ex}{\tiny$\substack{+0.7 \\ -0.1}$}}$\times10^{15}$   &  
 \phn$1.3${\raisebox{0.5ex}{\tiny$\substack{+0.1 \\ -0.3}$}}$\times10^{16}$ & \phn$500-1500$ & \phn11.7 & \phn0.1  \\   
Na$^{37}$Cl $v=1$    &   \phn$6.9${\raisebox{0.5ex}{\tiny$\substack{+1.3 \\ -0.2}$}}$\times10^{15}$   &  
 \phn$1.9${\raisebox{0.5ex}{\tiny$\substack{+0.1 \\ -0.9}$}}$\times10^{16}$ & \phn$500-1500$ & \phn11.7 & \phn0.1  \\   
Na$^{37}$Cl $v=5$    &  \phn$8.7${\raisebox{0.5ex}{\tiny$\substack{+1.4 \\ -0.9}$}}$\times10^{16}$   &  
 \phn$1.7${\raisebox{0.5ex}{\tiny$\substack{+0.2 \\ -0.9}$}}$\times10^{16}$ & \phn$500-1500$  & \phn11.7 & \phn0.1  \\   
Na$^{37}$Cl $v=7$    &  \phn$4.4${\raisebox{0.5ex}{\tiny$\substack{+1.4 \\ -0.9}$}}$\times10^{17}$   &   
 \phn$2.3${\raisebox{0.5ex}{\tiny$\substack{+0.2 \\ -1.1}$}}$\times10^{16}$ & \phn$500-1500$  & \phn11.7 & \phn0.1  \\   
KCl $v=0$     &   \phn$1.3${\raisebox{0.5ex}{\tiny$\substack{+0.1 \\ -0.3}$}}$\times10^{15}$   &  
 \phn$7.4${\raisebox{0.5ex}{\tiny$\substack{+0.3 \\ -1.3}$}}$\times10^{15}$ & \phn$500-1500$ & \phn10.3${\raisebox{0.5ex}{\tiny$\substack{+0.2 \\ -0.3}$}}$ & \phn0.3${\raisebox{0.5ex}{\tiny$\substack{+0.1 \\ -0.3}$}}$  \\
KCl $v=3$     &  \phn$5.1${\raisebox{0.5ex}{\tiny$\substack{+0.5 \\ -0.8}$}}$\times10^{15}$   &   
 \phn$6.0 ${\raisebox{0.5ex}{\tiny$\substack{+0.6 \\ -1.0}$}}$\times10^{15}$ & \phn$500-1500$ & \phn10.3 & \phn0.3   \\
KCl $v=4$     &   \phn$1.2${\raisebox{0.5ex}{\tiny$\substack{+0.1 \\ -0.5}$}}$\times10^{16}$  &   
\phn$8.2 ${\raisebox{0.5ex}{\tiny$\substack{+1.6 \\ -2.7}$}}$\times10^{15}$ & \phn$500-1500$ & \phn10.3 & \phn0.3   \\
KCl $v=5$     & \phn$1.8${\raisebox{0.5ex}{\tiny$\substack{+0.2 \\ -0.9}$}}$\times10^{16}$     &   
 \phn$7.2 ${\raisebox{0.5ex}{\tiny$\substack{+1.3 \\ -0.4}$}}$\times10^{15}$ & \phn$500-1500$ & \phn10.3 & \phn0.3   \\
KCl $v=6$     &  \phn$3.3${\raisebox{0.5ex}{\tiny$\substack{+0.7 \\ -0.5}$}}$\times10^{16}$    &   
 \phn$8.0 ${\raisebox{0.5ex}{\tiny$\substack{+1.2 \\ -3.2}$}}$\times10^{15}$ & \phn$500-1500$ & \phn10.3 & \phn0.3   \\
K$^{37}$Cl $v=1$     &   \phn$1.5${\raisebox{0.5ex}{\tiny$\substack{+0.1 \\ -0.5}$}}$\times10^{15}$    & 
 \phn$5.2${\raisebox{0.5ex}{\tiny$\substack{+0.2 \\ -0.2}$}}$\times10^{15}$  & \phn$500-1500$ & \phn10.3 & \phn0.3   \\
K$^{37}$Cl $v=4$     &  \phn$8.0${\raisebox{0.5ex}{\tiny$\substack{+0.6 \\ -0.6}$}}$\times10^{15}$  &  
 \phn$5.7${\raisebox{0.5ex}{\tiny$\substack{+0.3 \\ -0.7}$}}$\times10^{15}$  & \phn$500-1500$ & \phn10.3 & \phn0.3   \\
K$^{37}$Cl $v=6$     &   \phn$1.6${\raisebox{0.5ex}{\tiny$\substack{+0.1 \\ -0.6}$}}$\times10^{16}$ &  
 \phn$4.3${\raisebox{0.5ex}{\tiny$\substack{+0.5 \\ -0.4}$}}$\times10^{15}$ & \phn$500-1500$ & \phn10.3 & \phn0.3   \\
K$^{37}$Cl $v=7$     &  \phn$5.5${\raisebox{0.5ex}{\tiny$\substack{+4.5 \\ -3.7}$}}$\times10^{15}$ &  
 \phn$8.7${\raisebox{0.5ex}{\tiny$\substack{+9.1 \\ -7.1}$}}$\times10^{14}$ & \phn$500-1500$ & \phn10.3 & \phn0.3   \\
$^{41}$KCl $v=0$     &  \phn$5.6${\raisebox{0.5ex}{\tiny$\substack{+0.1\\ -0.1}$}}$\times10^{14}$   &  
 \phn$3.2${\raisebox{0.5ex}{\tiny$\substack{+0.1\\ -0.1}$}}$\times10^{15}$ & \phn$500-1500$ & \phn10.3 & \phn0.3   \\
$^{41}$KCl $v=2$     &   \phn$6.0${\raisebox{0.5ex}{\tiny$\substack{+1.1 \\ -0.2}$}}$\times10^{15}$   &  
{\phn$1.1${\raisebox{0.5ex}{\tiny$\substack{+0.1 \\ -0.6}$}}$\times10^{16}$} & \phn$500-1500$ & \phn10.3 & \phn0.3   \\
$^{41}$K$^{37}$Cl $v=1$  &  \phn$3.1${\raisebox{0.5ex}{\tiny$\substack{+0.4 \\ -0.4}$}}$\times10^{14}$   & 
 \phn$1.1${\raisebox{0.5ex}{\tiny$\substack{+0.1 \\ -0.1}$}}$\times10^{15}$ & \phn$500-1500$ & \phn10.3 & \phn0.3   \\
$^{41}$K$^{37}$Cl $v=2$  &    \phn$3.0${\raisebox{0.5ex}{\tiny$\substack{+0.6 \\ -0.6}$}}$\times10^{14}$  &  
 \phn$7.6${\raisebox{0.5ex}{\tiny$\substack{+1.5 \\ -1.3}$}}$\times10^{14}$ & \phn$500-1500$ & \phn10.3 & \phn0.3   \\
H$_2$O $v_2=1$ & &   \phn$3.0${\raisebox{0.5ex}{\tiny$\substack{+0.9 \\ -0.1}$}}$\times10^{20}$ & \multicolumn{1}{r}{$1500$} &  \phn19.1${\raisebox{0.5ex}{\tiny$\substack{+0.7 \\ -1.4}$}}$ & \phn-0.5${\raisebox{0.5ex}{\tiny$\substack{+0.1 \\ -0.1}$}}$ \\  
$^{29}$SiO $v=0$      &  &   \phn$5.9${\raisebox{0.5ex}{\tiny$\substack{+1.1 \\ -0.1}$}}$\times10^{17}$  & \multicolumn{1}{r}{$1500$} &  \phn23.5${\raisebox{0.5ex}{\tiny$\substack{+0.1 \\ -1.6}$}}$ & \phn-0.2${\raisebox{0.5ex}{\tiny$\substack{+0.4 \\ -0.1}$}}$  \\  
SiS $v=0$             & &  \phn$4.1${\raisebox{0.5ex}{\tiny$\substack{+0.1 \\ -0.1}$}}$\times10^{18}$ & \multicolumn{1}{r}{$1500$} & \phn18.6${\raisebox{0.5ex}{\tiny$\substack{+0.1 \\ -1.4}$}}$ & \phn-0.1${\raisebox{0.5ex}{\tiny$\substack{+0.3 \\ -0.1}$}}$  \\   
SiS $v=2$            &  &  $<$ 2.6$\times 10^{16}$ & \multicolumn{1}{r}{$1500$} & \phn18.6 & \phn-0.1  \\   

SiS $v=3$         &  &  \phn$2.8${\raisebox{0.5ex}{\tiny$\substack{+0.2 \\ -0.8}$}}$\times10^{14}$  & \multicolumn{1}{r}{$1500$} & \phn8.4${\raisebox{0.5ex}{\tiny$\substack{+0.9 \\ -0.4}$}}$ & \phn-1.8${\raisebox{0.5ex}{\tiny$\substack{+0.7 \\ -0.2}$}}$  \\   
Si$^{34}$S $v=0$      &  &   \phn$1.1${\raisebox{0.5ex}{\tiny$\substack{+0.1 \\ -0.1}$}}$\times10^{17}$  & \multicolumn{1}{r}{$1500$} & \phn16.7${\raisebox{0.5ex}{\tiny$\substack{+1.8 \\ -0.6}$}}$ & \phn0.1${\raisebox{0.5ex}{\tiny$\substack{+0.3 \\ -0.2}$}}$  \\  
Si$^{33}$S $v=0$      &  &   \phn$9.4${\raisebox{0.5ex}{\tiny$\substack{+0.6 \\ -0.6}$}}$\times10^{15}$  & \multicolumn{1}{r}{$1500$} & \phn15.2${\raisebox{0.5ex}{\tiny$\substack{+3.5 \\ -1.2}$}}$ & \phn-0.5${\raisebox{0.5ex}{\tiny$\substack{+0.9 \\ -1.0}$}}$  \\
SO$_2$ $v=0$      &  \multicolumn{2}{c}{ \phn$2.1${\raisebox{0.5ex}{\tiny$\substack{+0.1 \\ -0.1}$}}$\times10^{18}$} & \phn770${\raisebox{0.5ex}{\tiny$\substack{+120 \\ -370}$}}$ & \phn19.3${\raisebox{0.5ex}{\tiny$\substack{+1.9 \\ -3.0}$}}$ & \phn0.9${\raisebox{0.5ex}{\tiny$\substack{+1.0\\ -0.7}$}}$ \\
SO$_2$ $v_2=1$      &  \multicolumn{2}{c}{ \phn$7.3${\raisebox{0.5ex}{\tiny$\substack{+1.0 \\ -0.9}$}}$\times10^{17}$} & \phn770 & \phn19.3 & \phn0.9  \\ 
\tableline   
\end{tabular}
\tablecomments{Each vibrational level of the same species is fitted as an independent component, with its own partition function. The column density uncertainties do not include the calibration error.
Velocities are given relative to the NaCl $v=1 \ J=18-17$ centroid velocity used for the stacking procedure \citep[$v_{\rm LSR}=5.4~km~s^{-1}$;][]{Ginsburg2019}. For each molecule, the linewidth and offset velocity were fitted for one vibrational state and then fixed to those values when fitting the remaining vibrational states.} 
\end{center}

\end{table*}
%
\section{Additional line identifications}\label{sec:add-lines}
We used XCLASS to characterize the brightest lines in the observed spectrum previously reported by \citet{Ginsburg2019}, including unassigned features (U-lines) that we identify in this work. 
We first fitted the refractory compounds NaCl and KCl, which are well reproduced with rotational temperatures of $500-1500$~K. However, these temperatures differ from the values obtained by \citet{Ginsburg2019} using rotational diagrams: $T \approx 100$~K for KCl $v=0$ and NaCl $v=1$ and $v=2$, and $T > 1000$~K for other vibrationally excited states. Therefore, caution is needed concerning the column density estimates, since our data do not constrain the excitation temperature and non-LTE effects may also play a role, deserving further investigation in future work.

In addition, we searched for FeO, given the potential interest in comparing this species with FeS. However, we found all FeO transitions blended with intense lines (see Fig.~\ref{fig:total-spectrum}).

We also fitted H$_2$O $v_2=1$, Si$^{29}$O $v=0$ and SiS $v=0$. These molecular lines required rotational temperatures of $T_{\rm rot}\gtrsim1500$~K, as lower temperatures result in optically thick emission that fails to reproduce the observed intensities. SiO $v=0$ is affected by self-absorption. 
The previously unidentified features U214.637 and U232.634 reported by \citet{Ginsburg2019} are well reproduced by SiS $v=3$ and Si$^{33}$S $v=0$. 
Finally, we fitted SO$_2$, identifying U229.550 and U230.966 as $v_2=1$ and $v=0$ transitions, respectively. 
Since multiple transitions of SO$_2$ are detected in emission, we left the temperature free, obtaining a best-fit value of $\sim$770~K. 

The best-fit parameters for all fitted species are summarized in Table~\ref{tab:results}, and the synthetic spectrum is shown in Fig.~\ref{fig:total-spectrum}.  

%
\section{Discussion}\label{sec:discussion}

This paper presents the first search for gaseous FeS in the ISM. We have selected the massive disk Orion SrcI, where several molecules, including metal compounds, have previously been detected, and four ground-state transitions of FeS lie within the spectral bands covered by high-spatial-resolution ALMA observations. We report a tentative detection of FeS with a derived column density of $\rm 1.7\times10^{15}~cm^{-2}$, corresponding to an abundance of $\sim 9.4\times10^{-11}$ for an H$_2$ column density of $\rm 1.8\times10^{25}~cm^{-2}$. 

\begin{figure}[t]
    \centering
    \includegraphics[width=\linewidth]{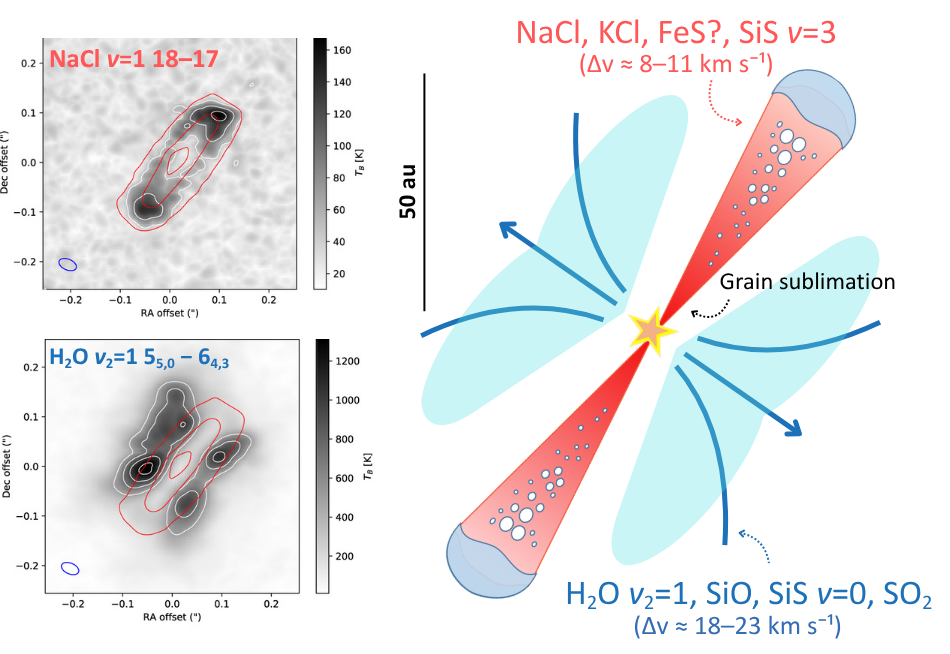}
    \caption{Illustration of the spatial distribution and proposed origin of the molecular species discussed in this work, including the tentative FeS emission. Left: H$_2$O $v_2=1$ and NaCl emission maps toward Orion SrcI reported by \citet{Ginsburg2018}. Right: schematic view of the proposed distinction between species associated with the outflow and those tracing the disk midplane, based on the results presented here (Table~\ref{tab:results}) and those of \citet{Ginsburg2018, Ginsburg2019}.}
    \label{fig:scheme}
\end{figure}

Our estimates show that gas-phase FeS is more than 10$^3$ times less abundant than SiS and SO$_2$ in Orion SrcI ($\rm 1.7\times10^{15}~cm^{-2}$ versus $ 4.1\times10^{18}$ and $\rm 2.1\times10^{18}~cm^{-2}$, respectively; see Table~\ref{tab:results}). However, the comparison with these sulfur-bearing species should be interpreted with caution, as they may trace different physical components along the line of sight. 
For example, maps of H$_2$O $v_2=1$ and SiO $v=0$ emission toward Orion SrcI show that these species primarily trace the outflow at high elevations above the disk midplane, whereas NaCl traces the disk surface and exhibits no coincident emission peaks with H$_2$O \citep{Ginsburg2018, Ginsburg2019}. Similarly, SiS $v=0$ mainly originates in the bipolar outflow, and SO$_2$ is associated with shocked, expanding gas surrounding the outflow \citep{Wright2020}. In this scenario, vibrationally excited SiS lines would be expected to be compact \citep[see e.g.,][]{Velilla2015} and might provide a more accurate comparison to the gas traced by metal salts, and possibly FeS. Figure~\ref{fig:scheme} provides a schematic view of the proposed spatial origin of the molecular species discussed here.
This interpretation is also supported with the different linewidth of the detected lines. The metal-bearing compounds and SiS $v=3$ present relatively narrow linewidths ($\sim$8--11~km~s$^{-1}$), whereas water, SiO, SO$_2$, and SiS $v=0$ lines exhibit broader profiles ($\sim$ 20~km~s$^{-1}$), consistent with part of their emission arising from the bipolar outflow. 
Moreover, the derived FeS column density is only about one order of magnitude lower than those obtained for the salts.
Comparing the SiS $v=3$ column density with FeS leads to SiS/FeS ratio of $\sim0.2-0.5$, slightly lower than the elemental Si/Fe ratio in dust \citep[$\sim$0.8; ][]{Voshchinnikov2010}. Considering the large uncertainties, this result is consistent with grain destruction playing an important role in the origin of the FeS emission. FeS may either be directly released through grain erosion or efficiently re-formed in the gas phase after refractory elements are liberated from dust grains.

It is also worth noting that the compounds SiS $v=0$ and SO$_2$ are the most abundant S-bearing molecules in our setup. 
In addition, SO$_2$ is predicted to be among the dominant sulfur-bearing species in hot-core chemical models \citep[e.g., ][]{Woods2015}. 
If we assume that most of the sulfur is contained in these molecules and 
considering that H$_2$O is expected to be one of the main oxygen carriers at high temperatures \citep{vanDishoeck2013}, we would obtain a first-order estimate of S/O $\approx$ 0.02.
This value is remarkably similar to that measured in the comet 67P \citep{Rubin2019} and a factor of two below the solar elemental ratio. Such a high sulfur abundance indicates that sulfur depletion is low in the warm inner disk where ices are sublimated. The similarity with cometary material suggests that some regions of the disk may exhibit comet-like volatile chemistry, potentially reflecting chemical inheritance between the protoplanetary disk and the bodies that eventually form within it.

%
\section{Summary and Conclusions}\label{sec:conclusions}
We present millimeter-wave spectroscopic calculations for FeS, which allow the first astronomical search for this species in the gas-phase ISM. 

We report a tentative detection of gaseous FeS using high-angular-resolution ALMA observations toward the massive disk Orion SrcI. We have identified two clean FeS ground-state transitions: FeS\,(19--18) $\Omega=4$ at 230.3836~GHz, and FeS\,(19--18) $\Omega=3$ at 230.9804~GHz. Further observations targeting additional transitions are needed to establish a robust detection of this species, since no archival high-angular-resolution data covering such transitions are available so far.
The best fit results in a column density of $1.7\times10^{15}$~cm$^{-2}$, which is three orders of magnitude lower than the most abundant S-bearing species SiS and SO$_2$.  

Our results are consistent with the chemically differentiated structure proposed by \citet{Ginsburg2019}, in which metallic salts arise from a thin layer of hot shocked gas where grains are destroyed. The similarity between the volatile S/O ratios derived here and those measured in comets suggests that some regions of the disk may exhibit comet-like volatile chemistry.

FeS may represent an important species in the hot gas associated with metallic salts. However, assessing its contribution to the refractory iron budget requires confirming the tentative FeS detection and better constraining its abundance. Future high-angular-resolution observations capable of resolving the different disk layers will be essential to disentangle the chemistry of the innermost regions, where refractory grains sublimate and metal-bearing molecules become abundant. Expanding the inventory of detected refractory species will provide direct constraints on dust composition, the mechanisms responsible for grain destruction, and the recycling of refractory elements into the gas phase, offering new insights into the chemical evolution of planet-forming disks.

\begin{acknowledgments}

This project has received funding from the European Research Council (ERC) under the European Union’s Horizon Europe research and innovation program ERC-AdG-2022 (GA No.\ 101096293). Funded by the European Union. Views and opinions expressed are however those of the author(s) only and do not necessarily reflect those of the European Union or the European Research Council Executive Agency. Neither the European Union nor the granting authority can be held responsible for them.
A.S.-M.\ acknowledges support from PID2023-146675NB-I00 (MCI-AEI-FEDER, UE) grant funded by MCIN/AEI/10.13039/501100011033 and by the European Union `Next GenerationEU’/PRTR, as well as the program Unidad de Excelencia María de Maeztu CEX2020-001058-M, and by the MaX-CSIC Excellence Award MaX4-SOMMA-IC.
A.F.\ acknowledge funding support from project PID2022-137980NB-I00 funded by the Spanish Ministry of Science and Innovation/State Agency of Research MCIN/AEI/10.13039/501100011033 and by `ERDF A way of making Europe'.
The work at Universt\"at zu K\"oln was supported by the Deutsche Forschungsgemeinschaft (DFG) via the collaborative research center SFB 1601 (project ID 500700252) subproject Inf (HSPM)
This paper makes use of the following ALMA data: . ALMA is a partnership of ESO (representing its member states), NSF (USA) and NINS (Japan), together with NRC (Canada), MOST and ASIAA (Taiwan), and KASI (Republic of Korea), in cooperation with the Republic of Chile. The Joint ALMA Observatory is operated by ESO, AUI/NRAO and NAOJ.
\end{acknowledgments}

\clearpage
\appendix
\restartappendixnumbering

\section{Supporting material: Tables and Figures}

\begin{table*}[h!]
\begin{center}
\caption{Spectroscopic properties of FeS transitions within the observed spectral coverage\label{tab:FeS}}
\begin{tabular}{lllcccc}
\tableline\tableline \noalign{\smallskip}
Line &  &  & Frequency & $\log_{10} I$ (300~K) & $E_\mathrm{u}$ & $g_\mathrm{u}$   \\ 
     &  &  & (MHz)  & (nm$^2$ MHz) & (K) &   \\ 
\tableline \noalign{\smallskip} 
FeS\, $v=0$ & $\Omega=4$ & $J=18^--17^+$ & 218267.9861(126)                     & -1.6332                    & 93.7113        & 37             \\
            &            & $J=18^+-17^-$ & 218267.9861(126)                     & -1.6332                    & 93.7113        & 37             \\
            &            & $J=19^--18^+$ & 230383.5728(123)                     & -1.5760                    & 104.7679       & 39             \\
            &            & $J=19^+-18^-$ & 230383.5728(123)                     & -1.5760                    & 104.7679       & 39             \\
            & $\Omega=3$ & $J=18^--17^+$ & 218833.5328(144)                     & -1.8315                    & 239.0237       & 37             \\
            &            & $J=18^+-17^-$ & 218833.5328(144)                     & -1.8315                    & 239.0237       & 37             \\
            &            & $J=19^--18^+$ & 230980.3707(140)                     & -1.7755                    & 250.1090       & 39             \\
            &            & $J=19^+-18^-$ & 230980.3707(140)                     & -1.7755                    & 250.1090       & 39             \\
FeS\, $v=1$ & $\Omega=4$ & $J=18^--17^+$ & 217250.0957(579)                     & -2.7155                    & 838.5609       & 37             \\
            &            & $J=18^+-17^-$ & 217250.0957(579)                     & -2.7155                    & 838.5609       & 37             \\
            &            & $J=19^--18^+$ & 229309.1141(573)                     & -2.6583               & 849.5660       & 39             \\
            &            & $J=19^+-18^-$ & 229309.1141(573)                     & -2.6583                    & 849.5660       & 39             \\
            & $\Omega=3$ & $J=18^--17^+$ & 217818.4424(561)                     & -2.9130                    & 983.2549       & 37             \\
            &            & $J=18^+-17^-$ & 217818.4424(561)                     & -2.9130                    & 983.2549       & 37             \\
            &            & $J=19^--18^+$ & 229908.8637(554)                     & -2.8569                    & 994.2887       & 39             \\
            &            & $J=19^+-18^-$ & 229908.8637(554)                     & -2.8569                    & 994.2887       & 39             \\
\tableline
\end{tabular}   
\end{center}
\end{table*}
\clearpage
\begin{figure*}
\centering
\includegraphics[width=\textwidth]{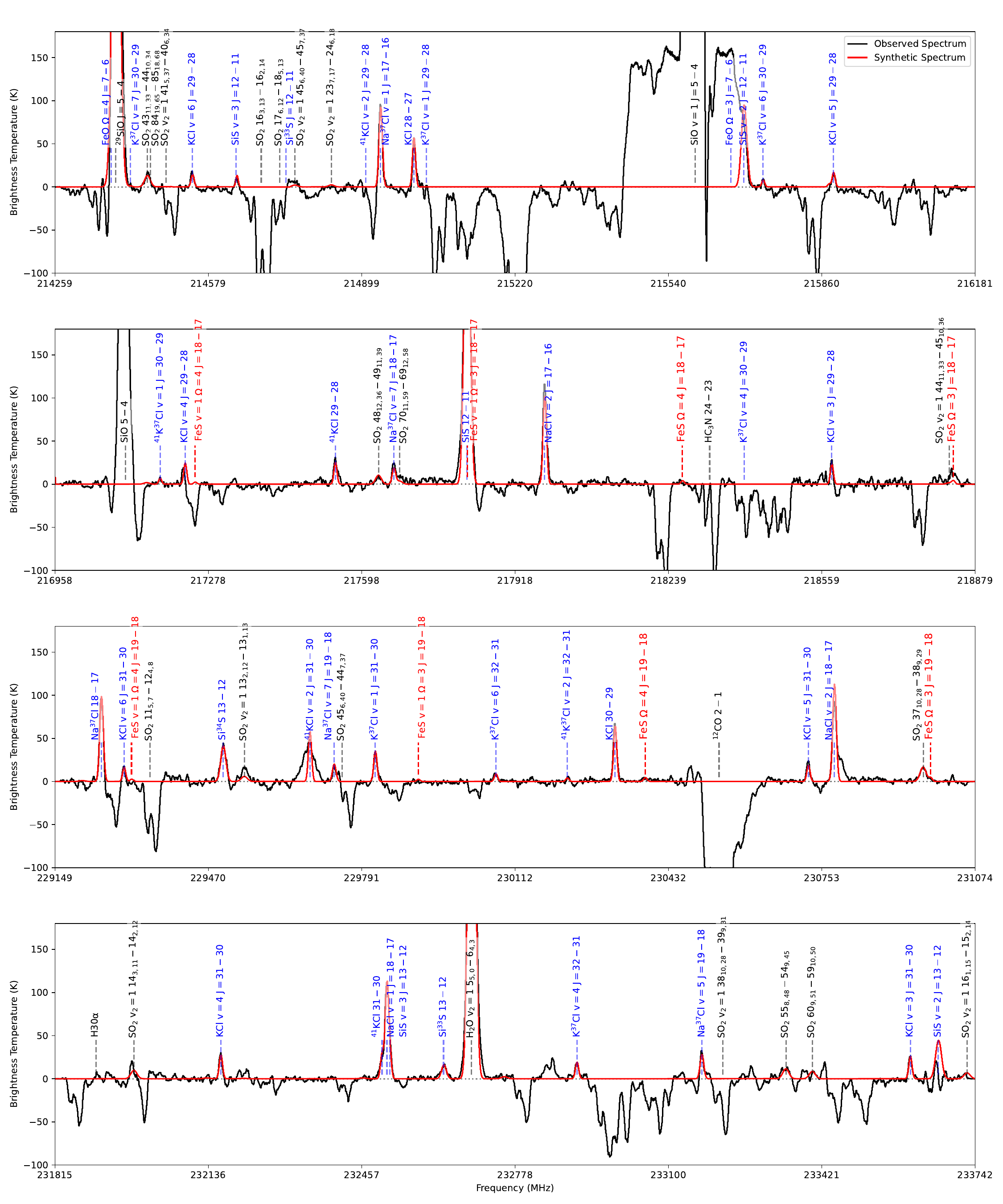}
\caption{Stacked spectrum from the Band 6 set-up toward Orion SrcI, with the intensity in brightness temperature. ALMA observational data are shown in black, while the synthetic spectrum, including the FeS contribution, is shown in red. Labeled transitions are listed in Table~\ref{tab:all_lines}. Frequencies of tentative FeS and blended FeO are also marked. Transitions affected by absorption were not considered in the fitting and are therefore not included in the synthetic spectrum.}
\label{fig:total-spectrum}
\end{figure*}
\clearpage

\startlongtable

\begin{deluxetable*}{lccccc}
\tablecaption{Summery of Spectroscopic Lines \label{tab:all_lines}}
\tablehead{ 
\colhead{Line}                                         & \colhead{Frequency} & \colhead{$\log_{10} A_\mathrm{ij}$} & \colhead{$E_\mathrm{u}$} & \colhead{$g_\mathrm{u}$} & \colhead{Ref.\tablenotemark{a}} \\
    &   \colhead{(MHz)}   & \colhead{(s$^{-1}$)} & \colhead{(K)} &   & 
    }
\startdata
FeO $\Omega=4$ $J=7-6$                  & 214376.3   & -2.84             & 26              & 15            &  (1)                  \\
$^{29}$SiO $J=5-4$                      & 214385.8   & -3.30             & 31              & 11            &  (2)                  \\
K$^{37}$Cl  $v=7$ $J=30-29$             & 214416.5   & -2.19             & 2846            & 61            &  (3)                 \\
SO$_2$ $43_{ 11,33}- 44_{ 10,34}$        & 214451.8   & -4.66             & 1166            & 87           &   (4)                   \\
SO$_2$ $84_{ 19,65}- 85_{ 18,68}$        & 214457.9   & -4.64             & 4155            & 169          &    (4)                 \\
SO$_2$ $v_2=1$  $41_{ 5,37}- 40_{ 6,34}$  & 214490.7 & 	-4.64	          & 1634	          & 83           &  (4)                \\
KCl $v=6$ $J=29-28$                   & 214545.8   & -2.19             & 2500            & 59              &     (3)              \\
SiS  $v=3$  $J=12-11$                    & 214637.2   & -3.78             & 3258            & 25           &    (5)                 \\
SO$_2$ $16_{3,13}-16_{2,14}$             & 214689.4   & -4.00             & 148             & 33           &     (4)                \\
SO$_2$ $17_{6,12}-18_{5,13}$             & 214728.3   & -4.72             & 229             & 35           &     (4)                \\
Si$^{33}$S $J=12-11$                    &   214741.0&   -3.78             &       67	&   25             &    (5)               \\
SO$_2$ $v_2=1$ $45_{ 6,40}- 45_{ 7,37}$  & 214759.8&  -4.59             	& 1826	      & 91              &     (4)         \\
SO$_2$ $v_2=1$ $23_{ 7,17}- 24_{ 6,18}$  & 214835.9   & -4.67             & 1143            & 47            &     (4)               \\
$^{41}$KCl $v=2$ $J=29-28$              & 214907.3   & -2.21             & 941             & 59             &    (3)                \\
Na$^{37}$Cl $v=1$ $J=17-16$              & 214938.7   & -2.34             & 607             & 35            &    (6)                \\
KCl $28-27$                            & 215008.3   & -2.28             & 150             & 57              &     (3)              \\
K$^{37}$Cl $v=1$ $J=29-28$               & 215034.6   & -2.22             & 548             & 59            &      (3)               \\
SiO $v=1$ $J=5-4$                      & 215596.0   & -3.29             & 1800            & 11              &      (2)             \\
FeO $\Omega=3$ $J=7-6$                 & 215669.8   & -2.79             & 299             & 15              &      (1)            \\
SiS $v=2$ $J=12-11$                    & 215697.5   & -3.77             & 2202            & 25              &     (5)             \\
K$^{37}$Cl $v=6$ $J=30-29$               & 215737.6   & -2.19             & 2473            & 61            &     (3)                \\
KCl $v=5$ $J=29-28$                    & 215884.7   & -2.19             & 2118            & 59              &      (3)             \\
SiO $5-4$                              & 217104.9   & -3.28             & 31              & 11              &         (2)          \\
$^{41}$K$^{37}$Cl $v=1$ $J=30-29$          & 217177.2   & -2.20             & 551             & 61          &       (3)                \\
KCl $v=4$ $J=29-28$                    & 217229.4   & -2.19             & 1733            & 59              &       (3)            \\
$^{41}$KCl $29-28$                       & 217543.2   & -2.21             & 157             & 59            &        (3)             \\
SO$_2$ $48_{12,36}-49_{11,39}$           & 217633.4   & -4.64             & 1433            & 97            &      (4)              \\
Na$^{37}$Cl $v=7$ $J=18-17$              & 217665.1   & -2.29             & 3596      & 37                  &      (5)             \\
SO$_2$ $70_{11,59}-69_{12,58}$           & 217677.2   & -4.60             & 2585            & 141           &       (4)             \\
SiS $12-11$                            & 217817.7   & -3.76             & 68              & 25              &      (5)            \\
NaCl $v=2$ $J=17-16$                   & 217980.2   & -2.31             & 1128            & 35              &      (5)            \\
HC$_3$N $24-23$                          & 218324.8   & -3.08             & 131             & 49            &       (7)            \\
K$^{37}$Cl $v=4$  $J=30-29$               &   218396.9  & -2.18	            & 1717	     & 61                 &      (3)             \\
KCl $v=3$ $J=29-28$                    & 218579.9   & -2.19             & 1345            & 59              &        (3)           \\
SO$_2$ $v_2=1$ $44_{ 11,33}- 45_{ 10,36}$ & 218825.5  &-4.59  &	1995 &	                 89                  &      (4)          \\
Na$^{37}$Cl $18-17$                      & 229246.1   & -2.26             & 105             & 37            &         (5)           \\
KCl $v=6$ $J=31-30$                    & 229293.6   & -2.11             & 2521            & 63               &        (3)          \\
SO$_2$ $11_{ 5, 7}-12_{ 4, 8}$           & 229347.6   & -4.72             & 122             & 23             &        (4)           \\
Si$^{34}$S $13-12$                       & 229500.9   & -3.69             & 77              & 27             &       (5)            \\
SO$_2$ $v_2=1$ $13_{ 2,12}- 13_{ 1,13}$  & 229545.3   & -4.15             & 854             & 27             &        (4)           \\
$^{41}$KCl $v=2$ $J=31-30$               & 229682.2   & -2.13             & 963             & 63             &        (3)            \\
Na$^{37}$Cl $v=7$ $J=19-18$              & 229732.8   & -2.21             & 3607            & 39             &         (5)          \\
SO$_2$ $45_{ 6,40}-44_{ 7,37}$           & 229749.7   & -4.55             & 1045            & 91             &        (4)           \\
K$^{37}$Cl $v=1$ $J=31-30$               & 229818.8   & -2.13             & 570             & 63             &          (3)          \\
K$^{37}$Cl $v=6$ $J=32-31$               & 230071.3   & -2.10             & 2495            & 65             &          (3)          \\
$^{41}$K$^{37}$Cl $v=2$ $J=32-31$          & 230220.7   & -2.12             & 957             & 65           &          (3)            \\
KCl $30-29$                            & 230320.6   & -2.20             & 171             & 61               &          (3)        \\
$^{12}$CO $2-1$                          & 230538.0   & -6.16             & 17              & 5              &          (8)         \\
KCl $v=5  (1)$ $J=31-30$                    & 230724.8   & -2.10             & 2140            & 63               &      (3)            \\
NaCl $v=2$ $J=18-17$                   & 230779.2   & -2.24             & 1139            & 37               &       (5)          \\
SO$_2$ $37_{10,28}-38_{ 9,29}$           & 230965.2   & -4.57             & 891             & 75             &     (4)              \\
H30$\alpha$                              & 231900.9   & $-$               & $-$             & $-$            &    (9)               \\
SO$_2$ $v_2=1$ $14_{ 3,11}- 14_{ 2,12}$  & 231980.5   & -3.93             & 881             & 29             &     (4)              \\
KCl $v=4$ $J=31-30$                    & 232162.2   & -2.10             & 1755            & 63               &          (3)        \\
$^{41}$KCl $31-30$                       & 232499.7   & -2.12             & 179             & 63             &          (3)          \\
NaCl $v=1$ $J=18-17$                   & 232510.0   & -2.23             & 626             & 37                &    (5)            \\
SiS $v=3$ $J=13-12$                    & 232515.8   & -3.67             & 3269            & 27                &     (5)           \\
Si$^{33}$S $13-12$                       & 232628.6   & -3.67             & 78              & 27             &      (5)             \\
H$_2$O $v_2 = 1$ $5_{5,0}-6_{4,3}$       & 232686.7   & -5.33             & 3462         & 33                &      (10)          \\
K$^{37}$Cl $v=4$ $J=32-31$               & 232907.9   & -2.10             & 1739            & 65             &        (3)            \\
Na$^{37}$Cl $v=5$ $J=19-18$              & 233169.2   & -2.21             & 2634            & 39             &     (5)              \\
SO$_2$ $v_2=1$ $38_{ 10,28}- 39_{ 9,31}$ & 233213.0   & -4.52             & 1708            & 77             &       (4)            \\
SO$_2$ $55_{ 8,48}- 54_{ 9,45}$          & 233345.5   & -4.51             & 1578            & 111            &       (4)            \\
SO$_2$ $60_{ 9,51}-59_{10,50}$           & 233400.9   & -4.51             & 1887            & 121            &       (4)            \\
KCl $v=3$ $J=31-30$                    & 233605.8   & -2.10             & 1367            & 63               &       (3)           \\
SiS $v=2$ $J=13-12$                    & 233664.4   & -3.67             & 2213            & 27               &       (5)          \\
SO$_2$ $v_2=1$ $16_{ 1,15}- 15_{ 2,14}$  & 233725.0   & -4.13             & 893             & 33             &        (4)          \\ 
\enddata
\tablenotetext{a}{References:  (1) \citet{FeO_rot_1996}, (2) \citet{SiO_rot_2013}, (3) \citet{KCl_rot_2004}, (4) \citet{SO2_rot_2005}, (5) \citet{SiS_rot_2007}, (6) \citet{NaCl_rot_2002}, (7) \citet{HC3N_rot_2000}, (8) \citet{CO_rot_1997}, (9) \citet{Halpha_1996}, (10) \citet{H2O_rot_2005}.} 
\end{deluxetable*}
\begin{table*}[h!]
\begin{center}
\caption{List of Potential Contaminants (within $\pm$4.2~MHz) of the Main FeS Transitions\label{tab:blending}}
\begin{tabular}{llcccl}
\tableline\tableline \noalign{\smallskip}
Line & Potential contaminants & $\log_{10} A_\mathrm{ij}$ & $E_\mathrm{low}$ & $g_\mathrm{u}$  & Ref.         \\ 
  &   & (s$^{-1}$) & (K) &  &   \\ 
\tableline \noalign{\smallskip}
FeS\,(19--18) $\Omega=4$            
     & H$_2$CNCN                                   & -2.85 & 161   & 45        &   \citet{H2CNCN_rot_1984}  \\
\phn at 230.3836~GHz
     & HC$_5$N $v_{11}=3$                                    & -2.88 & 933   & 170     &   \citet{HC5N_rot_2004}  \\
     & Si$^{33}$S $v=2$\tablenotemark{a}         & -3.08 & 2190   & 81    &   \citet{SiS_rot_2007}  \\
     & HNCHCN                                      & -3.26 & 522   & 49       &   \citet{HNCHCN_rot_1990}  \\
     & HNCHCN                                      & -3.26 & 522   & 49       &   \citet{HNCHCN_rot_1990}  \\
     & C$_3$H$_7$CN                                & -3.73 & 359   & 81    &   \citet{C3H7CN_rot_1988}  \\
     & C$_2$H$_5$CN $v_{12}=1$                              & -4.19 & 1240   & 91     &   \citet{C2H5CNv1_rot_2013}  \\
     & HOCHCHCHO                                   & -4.32 & 624   & 290     &   \citet{HOCHCHCHO_rot_2024}  \\
     & g-i-C$_3$H$_7$OH                           & -4.35 & 70   & 29      &   \citet{C3H7OH_rot_2006}  \\
     & C$_2$H$_5$OCHO                              & -4.36 & 1262  & 85      &   \citet{C2H5OCHO_rot_1967}  \\
     & C$_2$H$_5$OCHO                              & -4.36 & 1262  & 85      &   \citet{C2H5OCHO_rot_1967}   \\
     & NH$_2$CH$_2$CH$_2$OH $v_{25}=1,v_{27}=1$                       & -4.51 & 1112  & 87      &   \citet{NH2CH2CH2OH_rot_1971}  \\
     & C$_6$H$_5$CCH                               & -4.67 & 896   & 840       &   \citet{C6H5CCH_rot_2010}  \\
     & C$_6$H$_5$CCH                               & -4.67 & 896   & 840       &   \citet{C6H5CCH_rot_2010}  \\
     & $^{13}$CH$_3$CH$_2$CN                       & -4.89 & 486   & 77     &  \citet{13CH3CH2CN_rot_2007}  \\
     & $^{13}$CH$_3$CH$_2$CN                       & -4.89 & 486   & 77     &  \citet{13CH3CH2CN_rot_2007}  \\
     & CH$_3$O$^{13}$CHO $v_{18}=1$                          & -6.27 & 905   & 190       &  \citet{CH3O13CHO_rot_2010}  \\
     & CH$_3$COCH$_3$                              & -6.33 & 157   & 620     &   \citet{CH3COCH3_rot_2002}  \\
     & CH$_3$$^{13}$CHO                            & -6.62 & 726   & 79     &   \citet{CH313CHO_rot_2015}  \\
\tableline \noalign{\smallskip}
FeS\,(19--18) $\Omega=3$   
     & CH$_3$C$_3$N                                & -2.82 & 1380  & 230       &   \citet{CH3C3N_rot_1982}  \\
\phn at 230.9804~GHz
     & H$_2$$^{13}$CCCO                            & -2.88 & 88     & 33     &   \citet{H213CCCO_rot_2021}  \\
     & E-CH$_2$(CH)$_3$CN                          & -2.98 & 2060    & 490    &   \citet{E-C5H5CN_rot_2021}  \\
     & E-CH$_2$(CH)$_3$CN                          & -2.98 & 2060    & 490    &   \citet{E-C5H5CN_rot_2021}
     \\
     & c-C$^{13}$CH$_4$O                           & -3.80 & 1370  & 73        &   \citet{c-C13CH4O_rot_2022} \\
     & C$_3$H$_6$O$_2$                             & -4.57 & 1035  & 150     &   \citet{C3H6O2_rot_2010}  \\
     & CH$_3$CHOHCHO                               & -4.86 & 331   & 63      &   \citet{CH3CHOHCHO_rot_2019}  \\
     & CH$_3$CHOHCHO                               & -4.86 & 331   & 63      &  \citet{CH3CHOHCHO_rot_2019}  \\
     & p-c-SiC$_3$\tablenotemark{a}                & -4.93 & 581   & 89      &   \citet{p-c-SiC3_rot_2019}  \\ 
     & CH$_3$$^{13}$CH$_2$OH                       & -5.09 & 407   & 45       &   \citet{CH313CH2OH_rot_2012}  \\
     & CH$_3$$^{13}$CH$_2$OH                       & -5.09 & 407   & 45       &   \citet{CH313CH2OH_rot_2012}  \\
     & C$_2$H$_3$CN  $v_{11}=1$                              & -5.33 & 2448  & 190      &  \citet{C2H3CNv1_rot_2009}  \\
     & C$_2$H$_3$CN  $v_{11}=1$                              & -5.33 & 2448  & 190      &  \citet{C2H3CNv1_rot_2009}  \\
     & C$_2$H$_3$CN  $v_{11}=1$                              & -5.33 & 2448  & 190      & \citet{C2H3CNv1_rot_2009}  \\
     & CH$_3$OCHO                                  & -6.38 & 431   & 98     &  \citet{CH3OCHO_rot_2009} \\
     & CH$_3$OCHO                                  & -6.38 & 431   & 98     &  \citet{CH3OCHO_rot_2009} \\
     & CH$_3$COOH                                  & -6.51 & 1483 & 120         &   \citet{CH3COOH_rot_2013}  \\
     & CH$_3$COOH                                  & -6.51 & 1483 & 120         &   \citet{CH3COOH_rot_2013}  \\
  \tableline
\end{tabular}
\tablenotetext{a}{Refractory compounds that are potential contaminants.}
\end{center}
\end{table*}

\clearpage
\bibliography{refs}{}

@ARTICLE{Ginsburg2018,
       author = {{Ginsburg}, Adam and {Bally}, John and {Goddi}, Ciriaco and {Plambeck}, Richard and {Wright}, Melvyn},
        title = "{A Keplerian Disk around Orion SrCI, a {\ensuremath{\sim}} 15 M $_{{\ensuremath{\odot}}}$ YSO}",
      journal = {\apj},
         year = 2018,
        month = jun,
       volume = {860},
       number = {2},
          eid = {119},
        pages = {119},
          doi = {10.3847/1538-4357/aac205},
archivePrefix = {arXiv},
       eprint = {1804.10622},
 primaryClass = {astro-ph.GA},
       adsurl = {https://ui.adsabs.harvard.edu/abs/2018ApJ...860..119G}
}

@ARTICLE{Ginsburg2019,
       author = {{Ginsburg}, Adam and {McGuire}, Brett and {Plambeck}, Richard and {Bally}, John and {Goddi}, Ciriaco and {Wright}, Melvyn},
        title = "{Orion SrcI{\textquoteright}s Disk Is Salty}",
      journal = {\apj},
         year = 2019,
        month = feb,
       volume = {872},
       number = {1},
          eid = {54},
        pages = {54},
          doi = {10.3847/1538-4357/aafb71},
archivePrefix = {arXiv},
       eprint = {1901.04489},
 primaryClass = {astro-ph.GA},
       adsurl = {https://ui.adsabs.harvard.edu/abs/2019ApJ...872...54G}
}

@ARTICLE{Ginsburg2023,
       author = {{Ginsburg}, Adam and {McGuire}, Brett A. and {Sanhueza}, Patricio and {Olguin}, Fernando and {Maud}, Luke T. and {Tanaka}, Kei E.~I. and {Zhang}, Yichen and {Beuther}, Henrik and {Indriolo}, Nick},
        title = "{Salt-bearing Disk Candidates around High-mass Young Stellar Objects}",
      journal = {\apj},
         year = 2023,
        month = jan,
       volume = {942},
       number = {2},
          eid = {66},
        pages = {66},
          doi = {10.3847/1538-4357/ac9f4a},
archivePrefix = {arXiv},
       eprint = {2211.02502},
 primaryClass = {astro-ph.GA},
       adsurl = {https://ui.adsabs.harvard.edu/abs/2023ApJ...942...66G}
}

@ARTICLE{Liu2025,
       author = {{Liu}, Li},
        title = "{Formation mechanisms of pyrite in Earth's diverse geological systems}",
      journal = {Earth Science Reviews},
         year = 2025,
        month = nov,
       volume = {270},
          eid = {105234},
        pages = {105234},
          doi = {10.1016/j.earscirev.2025.105234},
       adsurl = {https://ui.adsabs.harvard.edu/abs/2025ESRv..27005234L}
}

@ARTICLE{Wachtershauser1988,
       author = {{W{\"a}chtersh{\"a}user}, G{\"u}nter},
        title = "{Pyrite Formation, the First Energy Source for Life: a Hypothesis}",
      journal = {Systematic and Applied Microbiology},
         year = 1988,
        month = aug,
       volume = {10},
       number = {3},
        pages = {207-210},
          doi = {10.1016/S0723-2020(88)80001-8},
       adsurl = {https://ui.adsabs.harvard.edu/abs/1988SyApM..10..207W}
}

@ARTICLE{Dwek2016,
       author = {{Dwek}, Eli},
        title = "{Iron: A Key Element for Understanding the Origin and Evolution of Interstellar Dust}",
      journal = {\apj},
         year = 2016,
        month = jul,
       volume = {825},
       number = {2},
          eid = {136},
        pages = {136},
          doi = {10.3847/0004-637X/825/2/136},
archivePrefix = {arXiv},
       eprint = {1605.01957},
 primaryClass = {astro-ph.GA},
       adsurl = {https://ui.adsabs.harvard.edu/abs/2016ApJ...825..136D}
}

@ARTICLE{Asplund2009,
       author = {{Asplund}, Martin and {Grevesse}, Nicolas and {Sauval}, A. Jacques and {Scott}, Pat},
        title = "{The Chemical Composition of the Sun}",
      journal = {\araa},
         year = 2009,
        month = sep,
       volume = {47},
       number = {1},
        pages = {481-522},
          doi = {10.1146/annurev.astro.46.060407.145222},
archivePrefix = {arXiv},
       eprint = {0909.0948},
 primaryClass = {astro-ph.SR},
       adsurl = {https://ui.adsabs.harvard.edu/abs/2009ARA&A..47..481A}
}

@ARTICLE{Jenkins2009,
       author = {{Jenkins}, Edward B.},
        title = "{A Unified Representation of Gas-Phase Element Depletions in the Interstellar Medium}",
      journal = {\apj},
         year = 2009,
        month = aug,
       volume = {700},
       number = {2},
        pages = {1299-1348},
          doi = {10.1088/0004-637X/700/2/1299},
archivePrefix = {arXiv},
       eprint = {0905.3173},
 primaryClass = {astro-ph.GA},
       adsurl = {https://ui.adsabs.harvard.edu/abs/2009ApJ...700.1299J}
}

@ARTICLE{Rogantini2026,
       author = {{Rogantini}, Daniele and {Canizares}, Claude and {Costantini}, Elisa and {Gu}, Liyi and {Mehdipour}, Missagh and {Psaradaki}, Ioanna and {Schulz}, Norbert S. and {Zeegers}, Sascha T.},
        title = "{Silicon, sulfur and iron in the interstellar medium: a high-resolution X-ray spectral study of GX 340+0}",
      journal = {\apj},
         year = 2026,
        month = apr,
       volume = {1002},
       number = {1},
        pages = {86},
          doi = {10.3847/1538-4357/ae5bb6},
archivePrefix = {arXiv},
       eprint = {2604.03375},
 primaryClass = {astro-ph.GA},
       adsurl = {https://ui.adsabs.harvard.edu/abs/2026arXiv260403375R},
}

@ARTICLE{Ruffle1999,
       author = {{Ruffle}, D.~P. and {Hartquist}, T.~W. and {Caselli}, P. and {Williams}, D.~A.},
        title = "{The sulphur depletion problem}",
      journal = {\mnras},
         year = 1999,
        month = jul,
       volume = {306},
       number = {3},
        pages = {691-695},
          doi = {10.1046/j.1365-8711.1999.02562.x},
       adsurl = {https://ui.adsabs.harvard.edu/abs/1999MNRAS.306..691R}
}

@ARTICLE{Fuente2023,
       author = {{Fuente}, A. and {Rivi{\`e}re-Marichalar}, P. and {Beitia-Antero}, L. and {Caselli}, P. and {Wakelam}, V. and {Esplugues}, G. and {Rodr{\'\i}guez-Baras}, M. and {Navarro-Almaida}, D. and {Gerin}, M. and {Kramer}, C. and {Bachiller}, R. and {Goicoechea}, J.~R. and {Jim{\'e}nez-Serra}, I. and {Loison}, J.~C. and {Ivlev}, A. and {Mart{\'\i}n-Dom{\'e}nech}, R. and {Spezzano}, S. and {Roncero}, O. and {Mu{\~n}oz-Caro}, G. and {Cazaux}, S. and {Marcelino}, N.},
        title = "{Gas phase Elemental abundances in Molecular cloudS (GEMS). VII. Sulfur elemental abundance}",
      journal = {\aap},
         year = 2023,
        month = feb,
       volume = {670},
          eid = {A114},
        pages = {A114},
          doi = {10.1051/0004-6361/202244843},
archivePrefix = {arXiv},
       eprint = {2212.03742},
 primaryClass = {astro-ph.GA},
       adsurl = {https://ui.adsabs.harvard.edu/abs/2023A&A...670A.114F}

}

@ARTICLE{Keller2002,
       author = {{Keller}, L.~P. and {Hony}, S. and {Bradley}, J.~P. and {Molster}, F.~J. and {Waters}, L.~B.~F.~M. and {Bouwman}, J. and {de Koter}, A. and {Brownlee}, D.~E. and {Flynn}, G.~J. and {Henning}, T. and {Mutschke}, H.},
        title = "{Identification of iron sulphide grains in protoplanetary disks}",
      journal = {\nat},
         year = 2002,
        month = may,
       volume = {417},
       number = {6884},
        pages = {148-150},
          doi = {10.1038/417148a},
       adsurl = {https://ui.adsabs.harvard.edu/abs/2002Natur.417..148K}
}

@ARTICLE{Tanaka2020,
       author = {{Tanaka}, Kei E.~I. and {Zhang}, Yichen and {Hirota}, Tomoya and {Sakai}, Nami and {Motogi}, Kazuhito and {Tomida}, Kengo and {Tan}, Jonathan C. and {Rosero}, Viviana and {Higuchi}, Aya E. and {Ohashi}, Satoshi and {Liu}, Mengyao and {Sugiyama}, Koichiro},
        title = "{Salt, Hot Water, and Silicon Compounds Tracing Massive Twin Disks}",
      journal = {\apjl},
         year = 2020,
        month = sep,
       volume = {900},
       number = {1},
          eid = {L2},
        pages = {L2},
          doi = {10.3847/2041-8213/abadfc},
archivePrefix = {arXiv},
       eprint = {2007.02962},
 primaryClass = {astro-ph.SR},
       adsurl = {https://ui.adsabs.harvard.edu/abs/2020ApJ...900L...2T}
}

@ARTICLE{Tasa-Chaveli2025,
       author = {{Tasa-Chaveli}, A. and {S{\'a}nchez-Monge}, {\'A}. and {Fuente}, A. and {Ginsburg}, A. and {M{\"u}ller}, H.~S.~P. and {M{\"o}ller}, T. and {Rivi{\`e}re-Marichalar}, P. and {Navarro-Almaida}, D. and {Esplugues}, G. and {Schilke}, P. and {Rodr{\'\i}guez-Baras}, M. and {Thorwirth}, S. and {Beitia-Antero}, L.},
        title = "{A Quest for Sulfur-bearing Refractory Species: Identification of CaS in the Interstellar Medium}",
      journal = {\apjl},
         year = 2025,
        month = nov,
       volume = {993},
       number = {2},
          eid = {L42},
        pages = {L42},
          doi = {10.3847/2041-8213/ae0e6b},
archivePrefix = {arXiv},
       eprint = {2510.02937},
 primaryClass = {astro-ph.GA},
       adsurl = {https://ui.adsabs.harvard.edu/abs/2025ApJ...993L..42T}
}

@ARTICLE{ReyMontejo2024,
       author = {{Rey-Montejo}, Marta and {Jim{\'e}nez-Serra}, Izaskun and {Mart{\'\i}n-Pintado}, Jes{\'u}s and {Rivilla}, V{\'\i}ctor M. and {Meg{\'\i}as}, Andr{\'e}s and {San Andr{\'e}s}, David and {Sanz-Novo}, Miguel and {Colzi}, Laura and {Zeng}, Shaoshan and {L{\'o}pez-Gallifa}, {\'A}lvaro and {Mart{\'\i}nez-Henares}, Antonio and {Mart{\'\i}n}, Sergio and {Tercero}, Bel{\'e}n and {de Vicente}, Pablo and {Requena-Torres}, Miguel},
        title = "{Discovery of MgS and NaS in the Interstellar Medium and Tentative Detection of CaO}",
      journal = {\apj},
         year = 2024,
        month = nov,
       volume = {975},
       number = {2},
          eid = {174},
        pages = {174},
          doi = {10.3847/1538-4357/ad736e},
archivePrefix = {arXiv},
       eprint = {2407.07693},
 primaryClass = {astro-ph.GA},
       adsurl = {https://ui.adsabs.harvard.edu/abs/2024ApJ...975..174R}
}

@ARTICLE{Pickett1998,
       author = {{Pickett}, H.~M. and {Poynter}, R.~L. and {Cohen}, E.~A. and {Delitsky}, M.~L. and {Pearson}, J.~C. and {M{\"u}ller}, H.~S.~P.},
        title = "{Submillimeter, millimeter and microwave spectral line catalog.}",
      journal = {\jqsrt},
         year = 1998,
        month = nov,
       volume = {60},
       number = {5},
        pages = {883-890},
          doi = {10.1016/S0022-4073(98)00091-0},
       adsurl = {https://ui.adsabs.harvard.edu/abs/1998JQSRT..60..883P}
}

@ARTICLE{Moeller2017,
       author = {{M{\"o}ller}, T. and {Endres}, C. and {Schilke}, P.},
        title = "{eXtended CASA Line Analysis Software Suite (XCLASS)}",
      journal = {\aap},
         year = 2017,
        month = feb,
       volume = {598},
          eid = {A7},
        pages = {A7},
          doi = {10.1051/0004-6361/201527203},
archivePrefix = {arXiv},
       eprint = {1508.04114},
 primaryClass = {astro-ph.IM},
       adsurl = {https://ui.adsabs.harvard.edu/abs/2017A&A...598A...7M}
}

@ARTICLE{Mueller2005,
       author = {{M{\"u}ller}, Holger S.~P. and {Schl{\"o}der}, Frank and {Stutzki}, J{\"u}rgen and {Winnewisser}, Gisbert},
        title = "{The Cologne Database for Molecular Spectroscopy, CDMS: a useful tool for astronomers and spectroscopists}",
      journal = {Journal of Molecular Structure},
         year = 2005,
        month = may,
       volume = {742},
       number = {1-3},
        pages = {215-227},
          doi = {10.1016/j.molstruc.2005.01.027},
       adsurl = {https://ui.adsabs.harvard.edu/abs/2005JMoSt.742..215M}
}

@ARTICLE{Apponi1999-c-SiC3,
       author = {{Apponi}, A.~J. and {McCarthy}, M.~C. and {Gottlieb}, C.~A. and {Thaddeus}, P.},
        title = "{Astronomical Detection of Rhomboidal SiC$_{3}$}",
      journal = {\apjl},
         year = 1999,
        month = may,
       volume = {516},
       number = {2},
        pages = {L103-L106},
          doi = {10.1086/311998},
       adsurl = {https://ui.adsabs.harvard.edu/abs/1999ApJ...516L.103A}
}

@ARTICLE{FeO_rot_1984,
       author = {{Endo}, Y. and {Saito}, S. and {Hirota}, E.},
        title = "{Laboratory millimeter-wave spectrum of iron monoxide, FeO}",
      journal = {\apjl},
         year = 1984,
        month = mar,
       volume = {278},
        pages = {L131},
          doi = {10.1086/184241},
       adsurl = {https://ui.adsabs.harvard.edu/abs/1984ApJ...278L.131E}
}

@ARTICLE{FeO_rot_1987,
       author = {{Kr{\"o}ckertskothen}, T. and {Kn{\"o}ckel}, H. and {Tiemann}, E.},
        title = "{Molecular beam rotational spectroscopy of the X$^{5}\Delta$ ground state of FeO}",
      journal = {Molecular Physics},
         year = 1987,
        month = jan,
       volume = {62},
       number = {4},
        pages = {1031-1040},
          doi = {10.1080/00268978700102761},
       adsurl = {https://ui.adsabs.harvard.edu/abs/1987MolPh..62.1031K}
}

@ARTICLE{FeO_rot_1996,
       author = {{Allen}, M.~D. and {Ziurys}, L.~M. and {Brown}, J.~M.},
        title = "{The millimeter-wave spectrum of FeO in its X$^{5}\Delta _{i}$ state ($v = 0$): a study of all five spin components}",
      journal = {Chemical Physics Letters},
         year = 1996,
        month = jul,
       volume = {257},
       number = {1},
        pages = {130-136},
          doi = {10.1016/0009-2614(96)00521-0},
       adsurl = {https://ui.adsabs.harvard.edu/abs/1996CPL...257..130A}
}

@ARTICLE{FeO_rot_2020,
       author = {{Wa{\ss}muth}, Bj{\"o}rn and {Breier}, Alexander A. and {Melosso}, Mattia and {Fuchs}, Guido W. and {Giesen}, Thomas F.},
        title = "{Rotational spectroscopy of rare iron monoxide isotopologues: A mass-independent analysis}",
      journal = {Molecular Physics},
         year = 2020,
        month = oct,
       volume = {118},
       number = {19-20},
          eid = {e1774087},
        pages = {e1774087},
          doi = {10.1080/00268976.2020.1774087},
       adsurl = {https://ui.adsabs.harvard.edu/abs/2020MolPh.11874087W}
}

@ARTICLE{FeS_rot_2004,
       author = {{Takano}, Shuro and {Yamamoto}, Satoshi and {Saito}, Shuji},
        title = "{The microwave spectrum of the FeS radical}",
      journal = {Journal of Molecular Spectroscopy},
         year = 2004,
        month = apr,
       volume = {224},
       number = {2},
        pages = {137-144},
          doi = {10.1016/j.jms.2004.01.003},
       adsurl = {https://ui.adsabs.harvard.edu/abs/2004JMoSp.224..137T}
}

@ARTICLE{FeS_vib_2011,
       author = {{Wang}, Li and {Huang}, Dao-ling and {Zhen}, Jun-feng and {Zhang}, Qun and {Chen}, Yang},
        title = "{Experimental Determination of the Vibrational Constants of FeS(X5{\ensuremath{\Delta}}) by Dispersed Fluorescence Spectroscopy}",
      journal = {Chinese Journal of Chemical Physics},
         year = 2011,
        month = feb,
       volume = {24},
       number = {1},
        pages = {1-3},
          doi = {10.1088/1674-0068/24/01/1-3},
       adsurl = {https://ui.adsabs.harvard.edu/abs/2011ChJCP..24....1W}
}

@ARTICLE{FeS_A-egy_1996,
       author = {{Zhang}, Nan and {Hayase}, Takasuke and {Kawamata}, Hiroshi and {Nakao}, Kojiro and {Nakajima}, Atsushi and {Kaya}, Koji},
        title = "{Photoelectron spectroscopy of iron-sulfur cluster anions}",
      journal = {\jcp},
         year = 1996,
        month = mar,
       volume = {104},
       number = {10},
        pages = {3413-3419},
          doi = {10.1063/1.471048},
       adsurl = {https://ui.adsabs.harvard.edu/abs/1996JChPh.104.3413Z}
}

@ARTICLE{FeS_MRCI_2022,
       author = {{da Silva}, Ramon S. and {Ballester}, Maikel Y.},
        title = "{Multireference configuration interaction calculations on the FeS molecule}",
      journal = {Molecular Physics},
         year = 2022,
        month = jul,
       volume = {120},
       number = {13},
          eid = {e2096142},
        pages = {e2096142},
          doi = {10.1080/00268976.2022.2096142},
       adsurl = {https://ui.adsabs.harvard.edu/abs/2022MolPh.12096142D}
}

@ARTICLE{spfit_1991,
       author = {{Pickett}, Herbert M.},
        title = "{The fitting and prediction of vibration-rotation spectra with spin interactions}",
      journal = {J. Mol. Spectrosc.},
         year = "1991",
        month = "Aug",
       volume = {148},
       number = {2},
        pages = {371-377},
          doi = {10.1016/0022-2852(91)90393-O},
       adsurl = {https://ui.adsabs.harvard.edu/abs/1991JMoSp.148..371P}
}

@ARTICLE{NO_rot_2015,
       author = {{M{\"u}ller}, Holger S.~P. and {Kobayashi}, Kaori and {Takahashi}, Kazumasa and {Tomaru}, Kazuko and {Matsushima}, Fusakazu},
        title = "{Terahertz spectroscopy of N$^{18}$O and isotopic invariant fit of several nitric oxide isotopologs}",
      journal = {Journal of Molecular Spectroscopy},
         year = 2015,
        month = apr,
       volume = {310},
        pages = {92-98},
          doi = {10.1016/j.jms.2014.12.002},
archivePrefix = {arXiv},
       eprint = {1412.4974},
 primaryClass = {astro-ph.GA},
       adsurl = {https://ui.adsabs.harvard.edu/abs/2015JMoSp.310...92M}
}

@ARTICLE{CH3O_rot_2025,
       author = {{Martin-Drumel}, Marie-Aline and {Spaniol}, Jean-Thibaut and {Chitarra}, Olivia and {Pirali}, Olivier and {M{\"u}ller}, Holger S.~P.},
        title = "{Submillimeter-Wave Spectroscopy of the CH$_3$O Radical}",
      journal = {Journal of Physical Chemistry A},
         year = 2025,
        month = jan,
       volume = {129},
       number = {3},
        pages = {754-761},
          doi = {10.1021/acs.jpca.4c07638},
archivePrefix = {arXiv},
       eprint = {2412.20990},
 primaryClass = {astro-ph.GA},
       adsurl = {https://ui.adsabs.harvard.edu/abs/2025JPCA..129..754M}
}

@ARTICLE{IO_rot_2001,
       author = {{Miller}, Charles E. and {Cohen}, Edward A.},
        title = "{Rotational spectroscopy of IO X $^{2}{\Pi}_{i}$}",
      journal = {J. Chem. Phys.},
         year = 2001,
        month = oct,
       volume = {115},
       number = {14},
        pages = {6459-6470},
          doi = {10.1063/1.1398308},
       adsurl = {https://ui.adsabs.harvard.edu/abs/2001JChPh.115.6459M}
}

@ARTICLE{SiS_rot_2007,
       author = {{M{\"u}ller}, H.~S.~P. and {McCarthy}, M.~C. and {Bizzocchi}, L. and {Gupta}, H. and {Esser}, S. and {Lichau}, H. and {Caris}, M. and {Lewen}, F. and {Hahn}, J. and {Degli Esposti}, C. and {Schlemmer}, S. and {Thaddeus}, P.},
        title = "{Rotational spectroscopy of the isotopic species of silicon monosulfide, SiS}",
      journal = {Physical Chemistry Chemical Physics},
         year = 2007,
        month = jan,
       volume = {9},
       number = {13},
        pages = {1579},
          doi = {10.1039/B618799D},
       adsurl = {https://ui.adsabs.harvard.edu/abs/2007PCCP....9.1579M}
}

@ARTICLE{SO2_rot_2005,
       author = {{M{\"u}ller}, Holger S.~P. and {Br{\"u}nken}, Sandra},
        title = "{Accurate rotational spectroscopy of sulfur dioxide, SO$_{2}$, in its ground vibrational and first excited bending states, $v_{2} = 0$, 1, up to 2~THz}",
      journal = {Journal of Molecular Spectroscopy},
         year = 2005,
        month = aug,
       volume = {232},
       number = {2},
        pages = {213-222},
          doi = {10.1016/j.jms.2005.04.010},
       adsurl = {https://ui.adsabs.harvard.edu/abs/2005JMoSp.232..213M}
}

@ARTICLE{Cook2017,
       author = {{Cook}, David L. and {Sch{\"o}nb{\"a}chler}, Maria},
        title = "{Iron Isotopic Compositions of Troilite (FeS) Inclusions from Iron Meteorites}",
      journal = {\aj},
         year = 2017,
        month = oct,
       volume = {154},
       number = {4},
          eid = {172},
        pages = {172},
          doi = {10.3847/1538-3881/aa8b12},
       adsurl = {https://ui.adsabs.harvard.edu/abs/2017AJ....154..172C}
}

@ARTICLE{Rubin2019,
       author = {{Rubin}, Martin and {Altwegg}, Kathrin and {Balsiger}, Hans and {Berthelier}, Jean-Jacques and {Combi}, Michael R. and {De Keyser}, Johan and {Drozdovskaya}, Maria and {Fiethe}, Bj{\"o}rn and {Fuselier}, Stephen A. and {Gasc}, S{\'e}bastien and {Gombosi}, Tamas I. and {H{\"a}nni}, Nora and {Hansen}, Kenneth C. and {Mall}, Urs and {R{\`e}me}, Henri and {Schroeder}, Isaac R.~H.~G. and {Schuhmann}, Markus and {S{\'e}mon}, Thierry and {Waite}, Jack H. and {Wampfler}, Susanne F. and {Wurz}, Peter},
        title = "{Elemental and molecular abundances in comet 67P/Churyumov-Gerasimenko}",
      journal = {\mnras},
         year = 2019,
        month = oct,
       volume = {489},
       number = {1},
        pages = {594-607},
          doi = {10.1093/mnras/stz2086},
archivePrefix = {arXiv},
       eprint = {1907.11044},
 primaryClass = {astro-ph.EP},
       adsurl = {https://ui.adsabs.harvard.edu/abs/2019MNRAS.489..594R}
}

@INCOLLECTION{Lodders2020,
       author = {{Lodders}, Katharina},
        title = "{Solar Elemental Abundances}",
    booktitle = {Oxford Research Encyclopedia of Planetary Science},
         year = 2020,
          eid = {145},
        pages = {145},
          doi = {10.1093/acrefore/9780190647926.013.145},
       adsurl = {https://ui.adsabs.harvard.edu/abs/2020oeps.book..145L}
}

@ARTICLE{Velilla2015,
       author = {{Velilla Prieto}, L. and {Cernicharo}, J. and {Quintana-Lacaci}, G. and {Ag{\'u}ndez}, M. and {Castro-Carrizo}, A. and {Fonfr{\'\i}a}, J.~P. and {Marcelino}, N. and {Z{\'u}{\~n}iga}, J. and {Requena}, A. and {Bastida}, A. and {Lique}, F. and {Gu{\'e}lin}, M.},
        title = "{Si-bearing Molecules Toward IRC+10216: ALMA Unveils the Molecular Envelope of CWLeo}",
      journal = {\apjl},
         year = 2015,
        month = jun,
       volume = {805},
       number = {2},
          eid = {L13},
        pages = {L13},
          doi = {10.1088/2041-8205/805/2/L13},
archivePrefix = {arXiv},
       eprint = {1504.05704},
 primaryClass = {astro-ph.SR},
       adsurl = {https://ui.adsabs.harvard.edu/abs/2015ApJ...805L..13V}
}

@ARTICLE{Wright2020,
       author = {{Wright}, Melvyn and {Plambeck}, Richard and {Hirota}, Tomoya and {Ginsburg}, Adam and {McGuire}, Brett and {Bally}, John and {Goddi}, Ciriaco},
        title = "{Observations of the Orion Source I Disk and Outflow Interface}",
      journal = {\apj},
         year = 2020,
        month = feb,
       volume = {889},
       number = {2},
          eid = {155},
        pages = {155},
          doi = {10.3847/1538-4357/ab5864},
archivePrefix = {arXiv},
       eprint = {1912.08862},
 primaryClass = {astro-ph.GA},
       adsurl = {https://ui.adsabs.harvard.edu/abs/2020ApJ...889..155W}
}

@ARTICLE{Thielemann1986,
       author = {{Thielemann}, F.-K. and {Nomoto}, K. and {Yokoi}, K.},
        title = "{Explosive nucleosynthesis in carbon deflagration models of Type I supernovae}",
      journal = {\aap},
         year = 1986,
        month = apr,
       volume = {158},
       number = {1-2},
        pages = {17-33},
       adsurl = {https://ui.adsabs.harvard.edu/abs/1986A&A...158...17T}
}

@ARTICLE{Kobayashi2011,
       author = {{Kobayashi}, Hiroshi and {Kimura}, Hiroshi and {Watanabe}, Sei-ichiro and {Yamamoto}, Tetsuo and {M{\"u}ller}, Sebastian},
        title = "{Sublimation temperature of circumstellar dust particles and its importance for dust ring formation}",
      journal = {Earth, Planets and Space},
         year = 2011,
        month = oct,
       volume = {63},
       number = {10},
        pages = {1067-1075},
          doi = {10.5047/eps.2011.03.012},
archivePrefix = {arXiv},
       eprint = {1104.5627},
 primaryClass = {astro-ph.EP},
       adsurl = {https://ui.adsabs.harvard.edu/abs/2011EP&S...63.1067K}
}

@ARTICLE{Woods2015,
       author = {{Woods}, Paul M. and {Occhiogrosso}, A. and {Viti}, S. and {Ka{\v{n}}uchov{\'a}}, Z. and {Palumbo}, M.~E. and {Price}, S.~D.},
        title = "{A new study of an old sink of sulphur in hot molecular cores: the sulphur residue}",
      journal = {\mnras},
         year = 2015,
        month = jun,
       volume = {450},
       number = {2},
        pages = {1256-1267},
          doi = {10.1093/mnras/stv652},
archivePrefix = {arXiv},
       eprint = {1503.07227},
 primaryClass = {astro-ph.SR},
       adsurl = {https://ui.adsabs.harvard.edu/abs/2015MNRAS.450.1256W}
}

@ARTICLE{vanDishoeck2013,
       author = {{van Dishoeck}, Ewine F. and {Herbst}, Eric and {Neufeld}, David A.},
        title = "{Interstellar Water Chemistry: From Laboratory to Observations}",
      journal = {Chemical Reviews},
         year = 2013,
        month = dec,
       volume = {113},
       number = {12},
        pages = {9043-9085},
          doi = {10.1021/cr4003177},
archivePrefix = {arXiv},
       eprint = {1312.4684},
 primaryClass = {astro-ph.GA},
       adsurl = {https://ui.adsabs.harvard.edu/abs/2013ChRv..113.9043V}
}

@ARTICLE{Voshchinnikov2010,
       author = {{Voshchinnikov}, N.~V. and {Henning}, Th.},
        title = "{From interstellar abundances to grain composition: the major dust constituents Mg, Si, and Fe}",
      journal = {\aap},
         year = 2010,
        month = jul,
       volume = {517},
          eid = {A45},
        pages = {A45},
          doi = {10.1051/0004-6361/200912817},
archivePrefix = {arXiv},
       eprint = {1004.0967},
 primaryClass = {astro-ph.GA},
       adsurl = {https://ui.adsabs.harvard.edu/abs/2010A&A...517A..45V}
}

@ARTICLE{Zhang2026-TiO,
       author = {{Zhang}, Ziwei E. and {Sakai}, Nami and {Agundez}, Marcelino and {Podio}, Linda and {Codella}, Claudio and {Hirota}, Tomoya and {Ginsburg}, Adam and {McGuire}, Brett and {Murillo}, Nadia M. and {Okoda}, Yuki and {Oya}, Yoko and {Sabatini}, Giovanni and {Takigawa}, Aki and {Tanaka}, Kei E.~I. and {Zhang}, Yichen and {Luo}, Qiuyi and {Cheng}, Yu and {Jimenez-Serra}, Izaskun and {Tachibana}, Shogo},
        title = "{Dust Seeding Molecules in a Massive Protostar -- Detection of TiO in Orion Source I}",
      journal = {arXiv e-prints},
         year = 2026,
        month = aug,
          eid = {arXiv:2608.01505},
        pages = {arXiv:2608.01505},
          doi = {10.48550/arXiv.2608.01505},
archivePrefix = {arXiv},
       eprint = {2608.01505},
 primaryClass = {astro-ph.GA},
       adsurl = {https://ui.adsabs.harvard.edu/abs/2026arXiv260801505Z}
}

@ARTICLE{SiO_rot_2013,
       author = {{M{\"u}ller}, Holger S.~P. and {Spezzano}, Silvia and {Bizzocchi}, Luca and {Gottlieb}, Carl A. and {Degli Esposti}, Claudio and {McCarthy}, Michael C.},
        title = "{Rotational Spectroscopy of Isotopologues of Silicon Monoxide, SiO, and Spectroscopic Parameters from a Combined Fit of Rotational and Rovibrational Data}",
      journal = {Journal of Physical Chemistry A},
         year = 2013,
        month = sep,
       volume = {117},
        pages = {13843-13854},
          doi = {10.1021/jp408391f},
       adsurl = {https://ui.adsabs.harvard.edu/abs/2013JPCA..11713843M}
}

@ARTICLE{NaCl_rot_2002,
       author = {{Caris}, M. and {Lewen}, F. and {Winnewisser}, G.},
        title = "{Pure Rotational Spectroscopy of Sodium Chloride, NaCl, up to 930 GHz}",
      journal = {Zeitschrift Naturforschung Teil A},
         year = 2002,
        month = aug,
       volume = {57},
       number = {8},
        pages = {663-668},
          doi = {10.1515/zna-2002-0805},
       adsurl = {https://ui.adsabs.harvard.edu/abs/2002ZNatA..57..663C}
}

@ARTICLE{KCl_rot_2004,
       author = {{Caris}, M. and {Lewen}, F. and {M{\"u}ller}, H.~S.~P. and {Winnewisser}, G.},
        title = "{Pure rotational spectroscopy of potassium chloride, KCl, up to 930GHz and isotopically invariant analysis of KCl and NaCl*1}",
      journal = {Journal of Molecular Structure},
         year = 2004,
        month = jun,
       volume = {695-696},
        pages = {243-251},
          doi = {10.1016/j.molstruc.2003.11.010},
       adsurl = {https://ui.adsabs.harvard.edu/abs/2004JMoSt.695..243C}
}

@ARTICLE{HC3N_rot_2000,
       author = {{Thorwirth}, S. and {M{\"u}ller}, H.~S.~P. and {Winnewisser}, G.},
        title = "{The Millimeter- and Submillimeter-Wave Spectrum of HC3N in the Ground and Vibrationally Excited States}",
      journal = {Journal of Molecular Spectroscopy},
         year = 2000,
        month = nov,
       volume = {204},
        pages = {133-144},
          doi = {10.1006/jmsp.2000.8209},
       adsurl = {https://ui.adsabs.harvard.edu/abs/2000JMoSp.204..133T}
}

@ARTICLE{CO_rot_1997,
       author = {{Winnewisser}, G. and {Belov}, S.~P. and {Klaus}, Th. and {Schieder}, R.},
        title = "{Sub-Doppler Measurements on the Rotational Transitions of Carbon Monoxide}",
      journal = {Journal of Molecular Spectroscopy},
         year = 1997,
        month = aug,
       volume = {184},
       number = {2},
        pages = {468-472},
          doi = {10.1006/jmsp.1997.7341},
       adsurl = {https://ui.adsabs.harvard.edu/abs/1997JMoSp.184..468W}
}

@ARTICLE{H2O_rot_2005,
       author = {{Pickett}, H.~M. and {Pearson}, J.~C. and {Miller}, C.~E.},
        title = "{Use of Euler series to fit spectra with application to water}",
      journal = {Journal of Molecular Spectroscopy},
         year = 2005,
        month = oct,
       volume = {233},
       number = {2},
        pages = {174-179},
          doi = {10.1016/j.jms.2005.06.013},
       adsurl = {https://ui.adsabs.harvard.edu/abs/2005JMoSp.233..174P}
}

@ARTICLE{HC5N_rot_2004,
       author = {{Yamada}, K.~M.~T. and {Degli Esposti}, C. and {Botschwina}, P. and {F{\"o}rster}, P. and {Bizzocchi}, L. and {Thorwirth}, S. and {Winnewisser}, G.},
        title = "{Millimeter-wave spectroscopy of HC$_{5}$N in vibrationally excited states below 500 cm$^{-1}$}",
      journal = {\aap},
         year = 2004,
        month = oct,
       volume = {425},
        pages = {767-772},
          doi = {10.1051/0004-6361:20040532},
       adsurl = {https://ui.adsabs.harvard.edu/abs/2004A&A...425..767Y}
}

@ARTICLE{H2CNCN_rot_1984,
       author = {{Winnewisser}, Manfred and {Winnewisser}, Brenda P. and {Wentrup}, Curt},
        title = "{The millimeter wave rotational spectrum of N-cyanomethanimine, CH $_{2}$NCN}",
      journal = {Journal of Molecular Spectroscopy},
         year = 1984,
        month = may,
       volume = {105},
       number = {1},
        pages = {193-205},
          doi = {10.1016/0022-2852(84)90111-5},
       adsurl = {https://ui.adsabs.harvard.edu/abs/1984JMoSp.105..193W}
}

@ARTICLE{HNCHCN_rot_1990,
       author = {{Takano}, Shuro and {Sugie}, Masaaki and {Sugawara}, Ko-ichi and {Takeo}, Harutoshi and {Matsumura}, Chi and {Masuda}, Akimasa and {Kuchitsu}, Kozo},
        title = "{Microwave spectra, dipole moments, and energy difference of E and Z C-cyanomethanimine (HCN dimer)}",
      journal = {Journal of Molecular Spectroscopy},
         year = 1990,
        month = may,
       volume = {141},
       number = {1},
        pages = {13-22},
          doi = {10.1016/0022-2852(90)90273-S},
       adsurl = {https://ui.adsabs.harvard.edu/abs/1990JMoSp.141...13T}
}

@ARTICLE{HOCHCHCHO_rot_2024,
       author = {{M{\"u}ller}, H.~S.~P. and {Coutens}, A. and {J{\o}rgensen}, J.~K. and {Margul{\`e}s}, L. and {Motiyenko}, R.~A. and {Guillemin}, J.-C.},
        title = "{The rotation-tunneling spectrum of 3-hydroxypropenal and confirmation of its detection toward IRAS 16293-2422 B}",
      journal = {\aap},
         year = 2024,
        month = jul,
       volume = {687},
          eid = {A8},
        pages = {A8},
          doi = {10.1051/0004-6361/202450029},
archivePrefix = {arXiv},
       eprint = {2405.01338},
 primaryClass = {astro-ph.GA},
       adsurl = {https://ui.adsabs.harvard.edu/abs/2024A&A...687A...8M}
}

@ARTICLE{C3H7OH_rot_2006,
       author = {{Maeda}, Atsuko and {Medvedev}, Ivan R. and {De Lucia}, Frank C. and {Herbst}, Eric},
        title = "{The Millimeter- and Submillimeter-Wave Spectrum of Iso-Propanol [(CH$_{3}$)$_{2}$CHOH]}",
      journal = {\apjs},
         year = 2006,
        month = oct,
       volume = {166},
       number = {2},
        pages = {650-658},
          doi = {10.1086/506595},
       adsurl = {https://ui.adsabs.harvard.edu/abs/2006ApJS..166..650M}
}

@ARTICLE{C2H5OCHO_rot_1967,
       author = {{Riveros}, Jos{\'e} M. and {Bright Wilson}, Jr., E.},
        title = "{Microwave Spectrum and Rotational Isomerism of Ethyl Formate}",
      journal = {\jcp},
         year = 1967,
        month = jun,
       volume = {46},
       number = {12},
        pages = {4605-4612},
          doi = {10.1063/1.1840610},
       adsurl = {https://ui.adsabs.harvard.edu/abs/1967JChPh..46.4605R}
}

@ARTICLE{NH2CH2CH2OH_rot_1971,
       author = {{Penn}, R.~E. and {Curl}, Jr., R.~F.},
        title = "{Microwave Spectrum of 2-Aminoethanol: Structural Effects of the Hydrogen Bond}",
      journal = {\jcp},
         year = 1971,
        month = jul,
       volume = {55},
       number = {2},
        pages = {651-658},
          doi = {10.1063/1.1676133},
       adsurl = {https://ui.adsabs.harvard.edu/abs/1971JChPh..55..651P}
}

@ARTICLE{C6H5CCH_rot_2010,
       author = {{Kisiel}, Zbigniew and {Kra{\'s}nicki}, Adam},
        title = "{The millimetre-wave rotational spectrum of phenylacetylene}",
      journal = {Journal of Molecular Spectroscopy},
         year = 2010,
        month = aug,
       volume = {262},
       number = {2},
        pages = {82-88},
          doi = {10.1016/j.jms.2010.05.007},
       adsurl = {https://ui.adsabs.harvard.edu/abs/2010JMoSp.262...82K}
}

@ARTICLE{13CH3CH2CN_rot_2007,
       author = {{Demyk}, K. and {M{\"a}der}, H. and {Tercero}, B. and {Cernicharo}, J. and {Demaison}, J. and {Margul{\`e}s}, L. and {Wegner}, M. and {Keipert}, S. and {Sheng}, M.},
        title = "{Isotopic ethyl cyanide $^{13}$CH3CH2CN, CH{}3$^{13}$CH2CN, and CH3CH{}2$^{13}$CN: laboratory rotational spectrum and detection in Orion}",
      journal = {\aap},
         year = 2007,
        month = apr,
       volume = {466},
       number = {1},
        pages = {255-259},
          doi = {10.1051/0004-6361:20066955},
       adsurl = {https://ui.adsabs.harvard.edu/abs/2007A&A...466..255D}
}

@ARTICLE{CH3O13CHO_rot_2010,
       author = {{Carvajal}, M. and {Kleiner}, I. and {Demaison}, J.},
        title = "{Global Assignment and Extension of Millimeter- and Submillimeter-wave Spectral Database of $^{13}$C$_{1}$-methyl Formate (H$^{13}$COOCH$_{3}$) in the Ground and First Excited States}",
      journal = {\apjs},
         year = 2010,
        month = oct,
       volume = {190},
       number = {2},
        pages = {315-321},
          doi = {10.1088/0067-0049/190/2/315},
       adsurl = {https://ui.adsabs.harvard.edu/abs/2010ApJS..190..315C}
}

@ARTICLE{CH3COCH3_rot_2002,
       author = {{Groner}, Peter and {Albert}, Sieghard and {Herbst}, Eric and {De Lucia}, Frank C. and {Lovas}, Frank J. and {Drouin}, Brian J. and {Pearson}, John C.},
        title = "{Acetone: Laboratory Assignments and Predictions through 620 GHZ for the Vibrational-Torsional Ground State}",
      journal = {\apjs},
         year = 2002,
        month = sep,
       volume = {142},
       number = {1},
        pages = {145-151},
          doi = {10.1086/341221},
       adsurl = {https://ui.adsabs.harvard.edu/abs/2002ApJS..142..145G}
}

@ARTICLE{CH313CHO_rot_2015,
       author = {{Margul{\`e}s}, L. and {Motiyenko}, R.~A. and {Ilyushin}, V.~V. and {Guillemin}, J.~C.},
        title = "{Millimeter and submillimeter wave spectra of mono-$^{13}$C-acetaldehydes}",
      journal = {\aap},
         year = 2015,
        month = jul,
       volume = {579},
          eid = {A46},
        pages = {A46},
          doi = {10.1051/0004-6361/201425478},
       adsurl = {https://ui.adsabs.harvard.edu/abs/2015A&A...579A..46M}
}

@ARTICLE{H213CCCO_rot_2021,
       author = {{M{\"u}ller}, Holger S.~P. and {Brahmi M.}, Ananya and {Guillemin}, Jean-Claude and {Lewen}, Frank and {Schlemmer}, Stephan},
        title = "{Rotational spectroscopy of isotopic cyclopropenone, c-H$_{2}$C$_{3}$O, and determination of its equilibrium structure}",
      journal = {\aap},
         year = 2021,
        month = mar,
       volume = {647},
          eid = {A179},
        pages = {A179},
          doi = {10.1051/0004-6361/202040088},
archivePrefix = {arXiv},
       eprint = {2102.03570},
 primaryClass = {astro-ph.IM},
       adsurl = {https://ui.adsabs.harvard.edu/abs/2021A&A...647A.179M}
}

@ARTICLE{E-C5H5CN_rot_2021,
       author = {{Zdanovskaia}, Maria A. and {Dorman}, P. Matisha and
                 {Orr}, Vanessa L. and {Owen}, Andrew N. and
                 {Kougias}, Samuel M. and {Esselman}, Brian J. and
                 {Woods}, R. Claude and {McMahon}, Robert J.},
        title = "{Rotational Spectra of Three Cyanobutadiene Isomers (C$_{5}$H$_{5}$N) of Relevance to Astrochemistry and Other Harsh Reaction Environments}",
      journal = {Journal of the American Chemical Society},
         year = 2021,
        month = jun,
       volume = {143},
       number = {25},
        pages = {9551-9564},
          doi = {10.1021/jacs.1c03777},
       adsurl = {}
}

@ARTICLE{c-C13CH4O_rot_2022,
       author = {{M{\"u}ller}, Holger S.~P. and {Guillemin}, Jean-Claude and {Lewen}, Frank and {Schlemmer}, Stephan},
        title = "{Rotational spectroscopy of isotopic oxirane, c-C$_{2}$H$_{4}$O}",
      journal = {Journal of Molecular Spectroscopy},
         year = 2022,
        month = feb,
       volume = {384},
          eid = {111584},
        pages = {111584},
          doi = {10.1016/j.jms.2022.111584},
archivePrefix = {arXiv},
       eprint = {2201.09266},
 primaryClass = {physics.chem-ph},
       adsurl = {https://ui.adsabs.harvard.edu/abs/2022JMoSp.38411584M}
}

@ARTICLE{C3H6O2_rot_2010,
       author = {{Braakman}, Rogier and {Drouin}, Brian J. and {Widicus Weaver}, Susanna L. and {Blake}, Geoffrey A.},
        title = "{Extended analysis of hydroxyacetone in the torsional ground state}",
      journal = {Journal of Molecular Spectroscopy},
         year = 2010,
        month = nov,
       volume = {264},
       number = {1},
        pages = {43-49},
          doi = {10.1016/j.jms.2010.09.003},
       adsurl = {https://ui.adsabs.harvard.edu/abs/2010JMoSp.264...43B}
}

@ARTICLE{CH3CHOHCHO_rot_2019,
       author = {{Alonso}, Elena R. and {McGuire}, Brett A. and {Kolesnikov{\'a}}, Lucie and {Carroll}, P. Brandon and {Le{\'o}n}, Iker and {Brogan}, Crystal L. and {Hunter}, Todd R. and {Guillemin}, Jean-Claude and {Alonso}, Jose L.},
        title = "{The Laboratory Millimeter and Submillimeter Rotational Spectrum of Lactaldehyde and an Astronomical Search in Sgr B2(N), Orion-KL, and NGC 6334I}",
      journal = {\apj},
         year = 2019,
        month = sep,
       volume = {883},
       number = {1},
          eid = {18},
        pages = {18},
          doi = {10.3847/1538-4357/ab3463},
       adsurl = {https://ui.adsabs.harvard.edu/abs/2019ApJ...883...18A}
}

@ARTICLE{p-c-SiC3_rot_2019,
       author = {{McCarthy}, Michael C. and {Gottlieb}, Carl A. and {Cernicharo}, Jose},
        title = "{Building blocks of dust: A coordinated laboratory and astronomical study of the archtype AGB carbon star IRC+10216}",
      journal = {Journal of Molecular Spectroscopy},
         year = 2019,
        month = feb,
       volume = {356},
        pages = {7-20},
          doi = {10.1016/j.jms.2018.11.018},
       adsurl = {https://ui.adsabs.harvard.edu/abs/2019JMoSp.356....7M}
}

@ARTICLE{CH313CH2OH_rot_2012,
       author = {{Bouchez}, Aurelia and {Walters}, Adam and {M{\"u}ller}, Holger S.~P. and {Ordu}, Matthias and {Lewen}, Frank and {Koerber}, Monika and {Bottinelli}, Sandrine and {Endres}, Christian P. and {Schlemmer}, Stephan},
        title = "{Millimetre-wave spectrum of anti-$^{13}$C$_{1}$ and $^{13}$C$_{2}$ isotopologues of ethanol}",
      journal = {\jqsrt},
         year = 2012,
        month = jul,
       volume = {113},
       number = {11},
        pages = {1148-1154},
          doi = {10.1016/j.jqsrt.2012.02.025},
       adsurl = {https://ui.adsabs.harvard.edu/abs/2012JQSRT.113.1148B}
}

@ARTICLE{C2H3CNv1_rot_2009,
       author = {{Kisiel}, Zbigniew and {Pszcz{\'o}{\l}kowski}, Lech and {Drouin}, Brian J. and {Brauer}, Carolyn S. and {Yu}, Shanshan and {Pearson}, John C.},
        title = "{The rotational spectrum of acrylonitrile up to 1.67 THz}",
      journal = {Journal of Molecular Spectroscopy},
         year = 2009,
        month = nov,
       volume = {258},
       number = {1-2},
        pages = {26-34},
          doi = {10.1016/j.jms.2009.08.011},
       adsurl = {https://ui.adsabs.harvard.edu/abs/2009JMoSp.258...26K}
}

@ARTICLE{CH3OCHO_rot_2009,
       author = {{Ilyushin}, V. and {Kryvda}, A. and {Alekseev}, E.},
        title = "{A new joint analysis of the ground and first excited torsional states of methylformate}",
      journal = {Journal of Molecular Spectroscopy},
         year = 2009,
        month = may,
       volume = {255},
       number = {1},
        pages = {32-38},
          doi = {10.1016/j.jms.2009.01.016},
       adsurl = {https://ui.adsabs.harvard.edu/abs/2009JMoSp.255...32I}
}

@ARTICLE{CH3COOH_rot_2013,
       author = {{Ilyushin}, Vadim V. and {Endres}, Christian P. and {Lewen}, Frank and {Schlemmer}, Stephan and {Drouin}, Brian J.},
        title = "{Submillimeter wave spectrum of acetic acid}",
      journal = {Journal of Molecular Spectroscopy},
         year = 2013,
        month = aug,
       volume = {290},
        pages = {31-41},
          doi = {10.1016/j.jms.2013.06.005},
       adsurl = {https://ui.adsabs.harvard.edu/abs/2013JMoSp.290...31I}
}

@ARTICLE{CH3C3N_rot_1982,
       author = {{Mo{\"\i}ses}, Alain and {Boucher}, Daniel and {Burie}, Jean and {Demaison}, Jean and {Dubrulle}, Alain},
        title = "{Millimeter-wave spectrum of methylcyanoacetylene}",
      journal = {Journal of Molecular Spectroscopy},
         year = 1982,
        month = apr,
       volume = {92},
       number = {2},
        pages = {497-498},
          doi = {10.1016/0022-2852(82)90118-7},
       adsurl = {https://ui.adsabs.harvard.edu/abs/1982JMoSp..92..497M}
}

@ARTICLE{C3H7CN_rot_1988,
       author = {{Wlodarczak}, G. and {Martinache}, L. and {Demaison}, J. and {Marstokk}, K.-M. and {M{\o}llendal}, Harald},
        title = "{Rotational spectrum of butyronitrile: Dipole moment, centrifugal distortion constants and energy difference between conformers}",
      journal = {Journal of Molecular Spectroscopy},
         year = 1988,
        month = jan,
       volume = {127},
       number = {1},
        pages = {178-185},
          doi = {10.1016/0022-2852(88)90017-3},
       adsurl = {https://ui.adsabs.harvard.edu/abs/1988JMoSp.127..178W}
}

@ARTICLE{C2H5CNv1_rot_2013,
       author = {{Daly}, A.~M. and {Berm{\'u}dez}, C. and {L{\'o}pez}, A. and {Tercero}, B. and {Pearson}, J.~C. and {Marcelino}, N. and {Alonso}, J.~L. and {Cernicharo}, J.},
        title = "{Laboratory Characterization and Astrophysical Detection of Vibrationally Excited States of Ethyl Cyanide}",
      journal = {\apj},
         year = 2013,
        month = may,
       volume = {768},
       number = {1},
          eid = {81},
        pages = {81},
          doi = {10.1088/0004-637X/768/1/81},
       adsurl = {https://ui.adsabs.harvard.edu/abs/2013ApJ...768...81D}
}

@ARTICLE{Halpha_1996,
       author = {{Towle}, Jonathan P. and {Feldman}, Paul A. and {Watson}, James K.~G.},
        title = "{A Catalog of Recombination Lines from 100 GHz to 10 Microns}",
      journal = {\apjs},
         year = 1996,
        month = dec,
       volume = {107},
        pages = {747},
          doi = {10.1086/192380},
       adsurl = {https://ui.adsabs.harvard.edu/abs/1996ApJS..107..747T}
}
\bibliographystyle{aasjournalv7}

\end{document}